\documentclass[longauth,twocolumn]{aa}
\usepackage{url,hyperref}
\makeatletter
\renewcommand*{\@fnsymbol}[1]{\ifcase#1\or*\or$\dagger$\or$\ddagger$\or**\or$\dagger\dagger$\or$\ddagger\ddagger$\fi}
\makeatother
\usepackage{graphicx}
\usepackage{txfonts}
\usepackage{xspace}
\usepackage{upgreek}
\newcommand{\suz}{\textit{Suzaku}\xspace}
\newcommand{\hess}{H.E.S.S.\xspace}
\newcommand{\nustar}{\textit{NuSTAR}\xspace}
\renewcommand{\arcmin}{\ensuremath{'}\xspace}
\renewcommand{\degr}{\ensuremath{^{\circ}}\xspace}

\begin{document} 

%\linenumbers

\title{\hess and \suz observations of the Vela~X pulsar wind nebula \protect\footnotemark[1]}

\authorrunning{H.E.S.S. Collaboration}
\titlerunning{\hess and \suz observations of the Vela~X pulsar wind nebula}

%%%%%%%%%% AUTHOR LIST %%%%%%%%%%%%%%%%%%%%%%

\author{H.E.S.S. Collaboration
\and H.~Abdalla \inst{\ref{NWU}}
\and F.~Aharonian \inst{\ref{MPIK},\ref{DIAS},\ref{RAU}}
\and F.~Ait~Benkhali \inst{\ref{MPIK}}
\and E.O.~Ang\"uner \inst{\ref{CPPM}}
\and M.~Arakawa \inst{\ref{Rikkyo}}
\and C.~Arcaro \inst{\ref{NWU}}
\and C.~Armand \inst{\ref{LAPP}}
\and M.~Backes \inst{\ref{UNAM},\ref{NWU}}
\and M.~Barnard \inst{\ref{NWU}}
\and Y.~Becherini \inst{\ref{Linnaeus}}
\and D.~Berge \inst{\ref{DESY}}
\and K.~Bernl\"ohr \inst{\ref{MPIK}}
\and R.~Blackwell \inst{\ref{Adelaide}}
\and M.~B\"ottcher \inst{\ref{NWU}}
\and C.~Boisson \inst{\ref{LUTH}}
\and J.~Bolmont \inst{\ref{LPNHE}}
\and S.~Bonnefoy \inst{\ref{DESY}}
\and J.~Bregeon \inst{\ref{LUPM}}
\and F.~Brun \inst{\ref{IRFU}}
\and P.~Brun \inst{\ref{IRFU}}
\and M.~Bryan \inst{\ref{GRAPPA}}
\and M.~B\"{u}chele \inst{\ref{ECAP}}
\and T.~Bulik \inst{\ref{UWarsaw}}
\and T.~Bylund \inst{\ref{Linnaeus}}
\and M.~Capasso \inst{\ref{IAAT}}
\and S.~Caroff \inst{\ref{LPNHE}} \protect\footnotemark[2]
\and A.~Carosi \inst{\ref{LAPP}}
\and S.~Casanova \inst{\ref{IFJPAN},\ref{MPIK}}
\and M.~Cerruti \inst{\ref{LPNHE},\ref{CerrutiNowAt}}
\and N.~Chakraborty \inst{\ref{MPIK}}
\and T.~Chand \inst{\ref{NWU}}
\and S.~Chandra \inst{\ref{NWU}}
\and R.C.G.~Chaves \inst{\ref{LUPM},\ref{CurieChaves}}
\and A.~Chen \inst{\ref{WITS}}
\and S.~Colafrancesco \inst{\ref{WITS}} \protect\footnotemark[3] 
\and B.~Condon \inst{\ref{CENB}}
\and I.D.~Davids \inst{\ref{UNAM}}
\and C.~Deil \inst{\ref{MPIK}}
\and J.~Devin \inst{\ref{LUPM}}
\and P.~deWilt \inst{\ref{Adelaide}}
\and L.~Dirson \inst{\ref{HH}}
\and A.~Djannati-Ata\"i \inst{\ref{APC}}
\and A.~Dmytriiev \inst{\ref{LUTH}}
\and A.~Donath \inst{\ref{MPIK}}
\and V.~Doroshenko \inst{\ref{IAAT}}
\and L.O'C.~Drury \inst{\ref{DIAS}}
\and J.~Dyks \inst{\ref{NCAC}}
\and K.~Egberts \inst{\ref{UP}}
\and G.~Emery \inst{\ref{LPNHE}}
\and J.-P.~Ernenwein \inst{\ref{CPPM}}
\and S.~Eschbach \inst{\ref{ECAP}}
\and K.~Feijen \inst{\ref{Adelaide}}
\and S.~Fegan \inst{\ref{LLR}}
\and A.~Fiasson \inst{\ref{LAPP}}
\and G.~Fontaine \inst{\ref{LLR}}
\and S.~Funk \inst{\ref{ECAP}}
\and M.~F\"u{\ss}ling \inst{\ref{DESY}}
\and S.~Gabici \inst{\ref{APC}}
\and Y.A.~Gallant \inst{\ref{LUPM}}
\and F.~Gat{\'e} \inst{\ref{LAPP}}
\and G.~Giavitto \inst{\ref{DESY}}
\and D.~Glawion \inst{\ref{LSW}}
\and J.F.~Glicenstein \inst{\ref{IRFU}}
\and D.~Gottschall \inst{\ref{IAAT}}
\and M.-H.~Grondin \inst{\ref{CENB}}
\and J.~Hahn \inst{\ref{MPIK}}
\and M.~Haupt \inst{\ref{DESY}}
\and G.~Heinzelmann \inst{\ref{HH}}
\and G.~Henri \inst{\ref{Grenoble}}
\and G.~Hermann \inst{\ref{MPIK}}
\and J.A.~Hinton \inst{\ref{MPIK}}
\and W.~Hofmann \inst{\ref{MPIK}}
\and C.~Hoischen \inst{\ref{UP}}
\and T.~L.~Holch \inst{\ref{HUB}}
\and M.~Holler \inst{\ref{LFUI}}
\and D.~Horns \inst{\ref{HH}}
\and D.~Huber \inst{\ref{LFUI}}
\and H.~Iwasaki \inst{\ref{Rikkyo}}
\and A.~Jacholkowska \inst{\ref{LPNHE}} \protect\footnotemark[3] 
\and M.~Jamrozy \inst{\ref{UJK}}
\and D.~Jankowsky \inst{\ref{ECAP}}
\and F.~Jankowsky \inst{\ref{LSW}}
\and L.~Jouvin \inst{\ref{APC}}
\and I.~Jung-Richardt \inst{\ref{ECAP}}
\and M.A.~Kastendieck \inst{\ref{HH}}
\and K.~Katarzy{\'n}ski \inst{\ref{NCUT}}
\and M.~Katsuragawa \inst{\ref{KAVLI}}
\and U.~Katz \inst{\ref{ECAP}}
\and D.~Khangulyan \inst{\ref{Rikkyo}} \protect\footnotemark[2]
\and B.~Kh\'elifi \inst{\ref{APC}}
\and J.~King \inst{\ref{LSW}}
\and S.~Klepser \inst{\ref{DESY}}
\and W.~Klu\'{z}niak \inst{\ref{NCAC}}
\and Nu.~Komin \inst{\ref{WITS}}
\and K.~Kosack \inst{\ref{IRFU}}
\and D.~Kostunin \inst{\ref{DESY}} 
\and M.~Kraus \inst{\ref{ECAP}}
\and G.~Lamanna \inst{\ref{LAPP}}
\and J.~Lau \inst{\ref{Adelaide}}
\and A.~Lemi\`ere \inst{\ref{APC}}
\and M.~Lemoine-Goumard \inst{\ref{CENB}}
\and J.-P.~Lenain \inst{\ref{LPNHE}}
\and E.~Leser \inst{\ref{UP},\ref{DESY}}
\and T.~Lohse \inst{\ref{HUB}}
\and R.~L\'opez-Coto \inst{\ref{MPIK}}
\and I.~Lypova \inst{\ref{DESY}}
\and D.~Malyshev \inst{\ref{IAAT}}
\and V.~Marandon \inst{\ref{MPIK}}
\and A.~Marcowith \inst{\ref{LUPM}}
\and C.~Mariaud \inst{\ref{LLR}}
\and G.~Mart\'i-Devesa \inst{\ref{LFUI}}
\and R.~Marx \inst{\ref{MPIK}}
\and G.~Maurin \inst{\ref{LAPP}}
\and N.I.~Maxted \inst{\ref{NSW}}
\and P.J.~Meintjes \inst{\ref{UFS}}
\and A.M.W.~Mitchell \inst{\ref{MPIK},\ref{MitchellNowAt}}
\and R.~Moderski \inst{\ref{NCAC}}
\and M.~Mohamed \inst{\ref{LSW}}
\and L.~Mohrmann \inst{\ref{ECAP}}
\and C.~Moore \inst{\ref{Leicester}}
\and E.~Moulin \inst{\ref{IRFU}}
\and T.~Murach \inst{\ref{DESY}}
\and S.~Nakashima  \inst{\ref{RIKKEN}}
\and M.~de~Naurois \inst{\ref{LLR}}
\and H.~Ndiyavala  \inst{\ref{NWU}}
\and F.~Niederwanger \inst{\ref{LFUI}}
\and J.~Niemiec \inst{\ref{IFJPAN}}
\and L.~Oakes \inst{\ref{HUB}}
\and P.~O'Brien \inst{\ref{Leicester}}
\and H.~Odaka \inst{\ref{Tokyo}} \protect\footnotemark[2]
\and S.~Ohm \inst{\ref{DESY}}
\and E.~de~Ona~Wilhelmi \inst{\ref{DESY}}
\and M.~Ostrowski \inst{\ref{UJK}}
\and I.~Oya \inst{\ref{DESY}}
\and M.~Panter \inst{\ref{MPIK}}
\and R.D.~Parsons \inst{\ref{MPIK}}
\and C.~Perennes \inst{\ref{LPNHE}}
\and P.-O.~Petrucci \inst{\ref{Grenoble}}
\and B.~Peyaud \inst{\ref{IRFU}}
\and Q.~Piel \inst{\ref{LAPP}}
\and S.~Pita \inst{\ref{APC}}
\and V.~Poireau \inst{\ref{LAPP}}
\and A.~Priyana~Noel \inst{\ref{UJK}}
\and D.A.~Prokhorov \inst{\ref{WITS}}
\and H.~Prokoph \inst{\ref{DESY}}
\and G.~P\"uhlhofer \inst{\ref{IAAT}}
\and M.~Punch \inst{\ref{APC},\ref{Linnaeus}}
\and A.~Quirrenbach \inst{\ref{LSW}}
\and S.~Raab \inst{\ref{ECAP}}
\and R.~Rauth \inst{\ref{LFUI}}
\and A.~Reimer \inst{\ref{LFUI}}
\and O.~Reimer \inst{\ref{LFUI}}
\and M.~Renaud \inst{\ref{LUPM}}
\and F.~Rieger \inst{\ref{MPIK}}
\and L.~Rinchiuso \inst{\ref{IRFU}}
\and C.~Romoli \inst{\ref{MPIK}}
\and G.~Rowell \inst{\ref{Adelaide}}
\and B.~Rudak \inst{\ref{NCAC}}
\and E.~Ruiz-Velasco \inst{\ref{MPIK}}
\and V.~Sahakian \inst{\ref{YPI}}
\and S.~Saito \inst{\ref{Rikkyo}}
\and D.A.~Sanchez \inst{\ref{LAPP}}
\and A.~Santangelo \inst{\ref{IAAT}}
\and M.~Sasaki \inst{\ref{ECAP}}
\and R.~Schlickeiser \inst{\ref{RUB}}
\and F.~Sch\"ussler \inst{\ref{IRFU}}
\and A.~Schulz \inst{\ref{DESY}}
\and H.~Schutte \inst{\ref{NWU}}
\and U.~Schwanke \inst{\ref{HUB}}
\and S.~Schwemmer \inst{\ref{LSW}}
\and M.~Seglar-Arroyo \inst{\ref{IRFU}}
\and M.~Senniappan \inst{\ref{Linnaeus}}
\and A.S.~Seyffert \inst{\ref{NWU}}
\and N.~Shafi \inst{\ref{WITS}}
\and I.~Shilon \inst{\ref{ECAP}}
\and K.~Shiningayamwe \inst{\ref{UNAM}}
\and R.~Simoni \inst{\ref{GRAPPA}}
\and A.~Sinha \inst{\ref{APC}}
\and H.~Sol \inst{\ref{LUTH}}
\and A.~Specovius \inst{\ref{ECAP}}
\and M.~Spir-Jacob \inst{\ref{APC}}
\and {\L.}~Stawarz \inst{\ref{UJK}}
\and R.~Steenkamp \inst{\ref{UNAM}}
\and C.~Stegmann \inst{\ref{UP},\ref{DESY}}
\and C.~Steppa \inst{\ref{UP}}
\and T.~Takahashi  \inst{\ref{KAVLI}}
\and J.-P.~Tavernet \inst{\ref{LPNHE}}
\and T.~Tavernier \inst{\ref{IRFU}}
\and A.M.~Taylor \inst{\ref{DESY}}
\and R.~Terrier \inst{\ref{APC}}
\and L. Tibaldo \inst{\ref{MPIK},\ref{TibaldoNowAt}} \protect\footnotemark[2]
\and D.~Tiziani \inst{\ref{ECAP}}
\and M.~Tluczykont \inst{\ref{HH}}
\and C.~Trichard \inst{\ref{LLR}}
\and M.~Tsirou \inst{\ref{LUPM}}
\and N.~Tsuji \inst{\ref{Rikkyo}}
\and R.~Tuffs \inst{\ref{MPIK}}
\and Y.~Uchiyama \inst{\ref{Rikkyo}}
\and D.J.~van~der~Walt \inst{\ref{NWU}}
\and C.~van~Eldik \inst{\ref{ECAP}}
\and C.~van~Rensburg \inst{\ref{NWU}}
\and B.~van~Soelen \inst{\ref{UFS}}
\and G.~Vasileiadis \inst{\ref{LUPM}}
\and J.~Veh \inst{\ref{ECAP}}
\and C.~Venter \inst{\ref{NWU}}
\and P.~Vincent \inst{\ref{LPNHE}}
\and J.~Vink \inst{\ref{GRAPPA}}
\and F.~Voisin \inst{\ref{Adelaide}}
\and H.J.~V\"olk \inst{\ref{MPIK}}
\and T.~Vuillaume \inst{\ref{LAPP}}
\and Z.~Wadiasingh \inst{\ref{NWU}}
\and S.J.~Wagner \inst{\ref{LSW}}
\and R.~White \inst{\ref{MPIK}}
\and A.~Wierzcholska \inst{\ref{IFJPAN}}
\and R.~Yang \inst{\ref{MPIK}}
\and H.~Yoneda \inst{\ref{KAVLI}}
\and D.~Zaborov \inst{\ref{LLR}}
\and M.~Zacharias \inst{\ref{NWU}}
\and R.~Zanin \inst{\ref{MPIK}}
\and A.A.~Zdziarski \inst{\ref{NCAC}}
\and A.~Zech \inst{\ref{LUTH}}
\and A.~Ziegler \inst{\ref{ECAP}}
\and J.~Zorn \inst{\ref{MPIK}}
\and N.~\.Zywucka \inst{\ref{NWU}}
}

\institute{
Centre for Space Research, North-West University, Potchefstroom 2520, South Africa \label{NWU} \and 
Universit\"at Hamburg, Institut f\"ur Experimentalphysik, Luruper Chaussee 149, D 22761 Hamburg, Germany \label{HH} \and 
Max-Planck-Institut f\"ur Kernphysik, P.O. Box 103980, D 69029 Heidelberg, Germany \label{MPIK} \and 
Dublin Institute for Advanced Studies, 31 Fitzwilliam Place, Dublin 2, Ireland \label{DIAS} \and 
% 5
High Energy Astrophysics Laboratory, RAU,  123 Hovsep Emin St  Yerevan 0051, Armenia \label{RAU} \and
Yerevan Physics Institute, 2 Alikhanian Brothers St., 375036 Yerevan, Armenia \label{YPI} \and
Institut f\"ur Physik, Humboldt-Universit\"at zu Berlin, Newtonstr. 15, D 12489 Berlin, Germany \label{HUB} \and
University of Namibia, Department of Physics, Private Bag 13301, Windhoek, Namibia, 12010 \label{UNAM} \and
GRAPPA, Anton Pannekoek Institute for Astronomy, University of Amsterdam,  Science Park 904, 1098 XH Amsterdam, The Netherlands \label{GRAPPA} \and
% 10
Department of Physics and Electrical Engineering, Linnaeus University,  351 95 V\"axj\"o, Sweden \label{Linnaeus} \and
Institut f\"ur Theoretische Physik, Lehrstuhl IV: Weltraum und Astrophysik, Ruhr-Universit\"at Bochum, D 44780 Bochum, Germany \label{RUB} \and
Institut f\"ur Astro- und Teilchenphysik, Leopold-Franzens-Universit\"at Innsbruck, A-6020 Innsbruck, Austria \label{LFUI} \and
School of Physical Sciences, University of Adelaide, Adelaide 5005, Australia \label{Adelaide} \and
% 15
LUTH, Observatoire de Paris, PSL Research University, CNRS, Universit\'e Paris Diderot, 5 Place Jules Janssen, 92190 Meudon, France \label{LUTH} \and
Sorbonne Universit\'e, Universit\'e Paris Diderot, Sorbonne Paris Cit\'e, CNRS/IN2P3, Laboratoire de Physique Nucl\'eaire et de Hautes Energies, LPNHE, 4 Place Jussieu, F-75252 Paris, France \label{LPNHE} \and
Laboratoire Univers et Particules de Montpellier, Universit\'e Montpellier, CNRS/IN2P3,  CC 72, Place Eug\`ene Bataillon, F-34095 Montpellier Cedex 5, France \label{LUPM} \and
IRFU, CEA, Universit\'e Paris-Saclay, F-91191 Gif-sur-Yvette, France \label{IRFU} \and
Astronomical Observatory, The University of Warsaw, Al. Ujazdowskie 4, 00-478 Warsaw, Poland \label{UWarsaw} \and
% 20
Aix Marseille Universit\'e, CNRS/IN2P3, CPPM, Marseille, France \label{CPPM} \and
Instytut Fizyki J\c{a}drowej PAN, ul. Radzikowskiego 152, 31-342 Krak{\'o}w, Poland \label{IFJPAN} \and
Funded by EU FP7 Marie Curie, grant agreement No. PIEF-GA-2012-332350 \label{CurieChaves}  \and
School of Physics, University of the Witwatersrand, 1 Jan Smuts Avenue, Braamfontein, Johannesburg, 2050 South Africa \label{WITS} \and
Laboratoire d'Annecy de Physique des Particules, Univ. Grenoble Alpes, Univ. Savoie Mont Blanc, CNRS, LAPP, 74000 Annecy, France \label{LAPP} \and
% 25
Landessternwarte, Universit\"at Heidelberg, K\"onigstuhl, D 69117 Heidelberg, Germany \label{LSW} \and
Universit\'e Bordeaux, CNRS/IN2P3, Centre d'\'Etudes Nucl\'eaires de Bordeaux Gradignan, 33175 Gradignan, France \label{CENB} \and
Oskar Klein Centre, Department of Physics, Stockholm University, Albanova University Center, SE-10691 Stockholm, Sweden \label{OKC} \and
Institut f\"ur Astronomie und Astrophysik, Universit\"at T\"ubingen, Sand 1, D 72076 T\"ubingen, Germany \label{IAAT} \and
Laboratoire Leprince-Ringuet, Ecole Polytechnique, CNRS/IN2P3, F-91128 Palaiseau, France \label{LLR} \and
% 30
APC, AstroParticule et Cosmologie, Universit\'{e} Paris Diderot, CNRS/IN2P3, CEA/Irfu, Observatoire de Paris, Sorbonne Paris Cit\'{e}, 10, rue Alice Domon et L\'{e}onie Duquet, 75205 Paris Cedex 13, France \label{APC} \and
Univ. Grenoble Alpes, CNRS, IPAG, F-38000 Grenoble, France \label{Grenoble} \and
Department of Physics and Astronomy, The University of Leicester, University Road, Leicester, LE1 7RH, United Kingdom \label{Leicester} \and
Nicolaus Copernicus Astronomical Center, Polish Academy of Sciences, ul. Bartycka 18, 00-716 Warsaw, Poland \label{NCAC} \and
Institut f\"ur Physik und Astronomie, Universit\"at Potsdam,  Karl-Liebknecht-Strasse 24/25, D 14476 Potsdam, Germany \label{UP} \and
% 35
Friedrich-Alexander-Universit\"at Erlangen-N\"urnberg, Erlangen Centre for Astroparticle Physics, Erwin-Rommel-Str. 1, D 91058 Erlangen, Germany \label{ECAP} \and
DESY, D-15738 Zeuthen, Germany \label{DESY} \and
Obserwatorium Astronomiczne, Uniwersytet Jagiello{\'n}ski, ul. Orla 171, 30-244 Krak{\'o}w, Poland \label{UJK} \and
Centre for Astronomy, Faculty of Physics, Astronomy and Informatics, Nicolaus Copernicus University,  Grudziadzka 5, 87-100 Torun, Poland \label{NCUT} \and
Department of Physics, University of the Free State,  PO Box 339, Bloemfontein 9300, South Africa \label{UFS} \and
% 40
Department of Physics, Rikkyo University, 3-34-1 Nishi-Ikebukuro, Toshima-ku, Tokyo 171-8501, Japan \label{Rikkyo} \and
Kavli Institute for the Physics and Mathematics of the Universe (Kavli IPMU), The University of Tokyo Institutes for Advanced Study (UTIAS), The University of Tokyo, 5-1-5 Kashiwa-no-Ha, Kashiwa City, Chiba, 277-8583, Japan \label{KAVLI} \and
Department of Physics, The University of Tokyo, 7-3-1 Hongo, Bunkyo-ku, Tokyo 113-0033, Japan \label{Tokyo} \and
RIKEN, 2-1 Hirosawa, Wako, Saitama 351-0198, Japan \label{RIKKEN} \and
% 45
% Nigel's affiliation
School of Science, The University of New South Wales, Australian Defence Force Academy, Canberra 2610, Australia \label{NSW} \and
%% Affiliation of people who left the collaboration
Now at Institut de Ci\`{e}ncies del Cosmos (ICC UB), Universitat de Barcelona (IEEC-UB), Mart\'{i} Franqu\`es 1, E08028 Barcelona, Spain \label{CerrutiNowAt} \and
Now at Physik Institut, Universit\"at Z\"urich, Winterthurerstrasse 190, CH-8057 Z\"urich, Switzerland \label{MitchellNowAt} \and
Now at IRAP, Universit\'e de Toulouse, CNRS, UPS, CNES, 9 avenue Colonel-Roche, 31028 Toulouse, Cedex 4, France \label{TibaldoNowAt}
}

\offprints{H.E.S.S.~collaboration,
\protect\\\email{\href{mailto:contact.hess@hess-experiment.eu}{contact.hess@hess-experiment.eu}};
\protect\\\protect\footnotemark[1] Spectra are only available in electronic form at the CDS via anonymous ftp to cdsarc.u-strasbg.fr (130.79.128.5) or via \url{http://cdsweb.u-strasbg.fr/cgi-bin/qcat?J/A+A/}
\protect\\\protect\footnotemark[2] Corresponding authors
\protect\\\protect\footnotemark[3] Deceased
}

%%%%%%%%%% AUTHOR LIST %%%%%%%%%%%%%%%%%%%%%%

  \abstract
  { Pulsar wind nebulae (PWNe) represent the most prominent population of Galactic very-high-energy gamma-ray sources and are
    thought to be an efficient source of leptonic cosmic rays. Vela~X is a nearby middle-aged PWN, which shows bright X-ray and TeV gamma-ray emission towards an elongated structure called the cocoon.
    } 
     {Since TeV emission is likely inverse-Compton emission of electrons, predominantly from interactions with the cosmic microwave background, while X-ray emission is synchrotron radiation of the same electrons, we aim to derive the properties of the relativistic particles and of magnetic fields with minimal modelling.} 
    {We used data from the \suz~XIS to derive the spectra from three compact regions in Vela X
    covering distances from \(0.3\rm\,pc\) to \(4\rm\,pc\) from the pulsar along the cocoon. We obtained gamma-ray spectra of the same regions from \hess observations and fitted a radiative model to the multi-wavelength spectra.} 
     {The TeV electron spectra and magnetic field strengths are consistent within the uncertainties for the three regions, with energy densities of the order $10^{-12}\rm\,erg\,cm^{-3}$. The data indicate the presence of a cutoff in the electron spectrum at energies of $\sim$$100\rm\,TeV$ and a magnetic field strength of $\sim$$6\,\rm\upmu G$. Constraints on the presence of turbulent magnetic fields are weak.}
   {The pressure of TeV electrons and magnetic fields in the cocoon is  dynamically negligible, requiring the presence of another dominant pressure component to balance the pulsar wind at the termination shock. Sub-TeV electrons cannot completely account for the missing pressure, which may be provided either by relativistic ions or from mixing of the ejecta with the pulsar wind. 
The electron spectra are consistent with expectations from transport scenarios dominated either by
advection via the reverse shock or by diffusion, but for the latter the role of radiative losses near the termination shock needs to be further investigated in the light of the measured cutoff energies.
Constraints on turbulent magnetic fields and the shape of the electron cutoff can be improved by spectral measurements in the energy range $\gtrsim 10\rm\,keV$.}

   \keywords{stars: winds, outflows --
                         gamma rays: stars --
             pulsars: individual PSR~B0833$-$45 (Vela pulsar) --
             acceleration of particles --
             radiation mechanisms: non-thermal
               }

\maketitle
%
%-------------------------------------------------------------------

\section{Introduction}

Pulsars eject relativistic winds that are thought to be loaded primarily with electrons and positrons. The wind is
highly supersonic, leading to the formation of a termination shock at the distance where the wind ram pressure becomes
comparable to the external pressure. Beyond the shock lies a bubble of magnetised relativistic plasma that originated in
the pulsar magnetosphere. The formation of these so-called pulsar wind nebulae (PWNe) is accompanied by efficient
particle acceleration. Thus, PWNe are bright non-thermal emitters with spectra extending from radio to gamma rays, and
represent the dominant class of identified Galactic sources observed at the highest-energy end of the electromagnetic spectrum
\citep{hess2017pwne,hgps2018}. However, the exact sites and mechanisms of particle acceleration up to PeV energies in PWNe, and
to what extent the PWNe contribute to the electron and positron component observed in cosmic rays still remain to be established
\citep[e.g.][]{amato2014}.

Most of the recent progress in understanding PWNe, including acceleration and propagation of high-energy particles, has come from theoretical studies of the properties of the magnetohydrodynamic (MHD) flow
in the nebula \citep{KenCor84,BogKha02,Lyub02,2006A&A...453..621D,2008A&A...485..337V,amato2014,pkk14}, and observations in the X-ray and TeV gamma-ray
domains
\citep[for a review see, e.g.][]{kargaltsev2015}.
Under the conditions inferred in PWNe, synchrotron radiation of relativistic electrons should dominate the emission at X-ray energies, and the TeV gamma rays are generated through inverse-Compton (IC) scattering. 
For typical magnetic fields in PWNe, {X-ray and TeV gamma-ray emissions are} generated by particles with similar energies. The synchrotron emissivity is determined by the electron
density and the strength of the local magnetic field. MHD simulations show that the magnetic field
strength can vary strongly throughout the nebula. Thus, X-ray data alone do not enable us to determine the particle density or to perform detailed studies of the particle evolution in PWNe.

On the other hand, the IC emissivity is determined by the electron density and the energy density of target photon fields. For the production of TeV emission, the latter are predominantly the cosmic microwave background (CMB), and far-infrared (FIR) Galactic dust emission, which are unrelated to the nebular processes, contrary to magnetic fields that determine the synchrotron emission. This means that gamma rays are a direct tracer of the particle densities in a nearly model-independent way (except for uncertainties related to the FIR photon field). Once the
particle density is estimated, the X-ray emissivity constrains the strength of the magnetic field. Therefore, in principle, the combination of X-rays and gamma~rays can
provide invaluable information to study particle acceleration and transport in PWNe, and validate numerical
MHD simulations.

However, combining information from the X-ray and TeV wavebands is often complicated, not only because of the unavoidable
projection effects along the line of sight, but also due to the limited angular resolution of gamma-ray
measurements. TeV signals are typically registered from large structures that have varying magnetic field strengths and particles accumulated over a long fraction of the
pulsar's lifetime \citep[e.g.][]{dejager2009}.

The latter limitation can be overcome with observations for a sufficiently extended and bright PWN, such as Vela X.
Originally discovered as a $3\degr \times 2\degr$ radio source in the Vela constellation \citep{rishbeth1958}, owing to its level of polarisation and its flat spectrum Vela X was later interpreted by \citet{weiler1980} as the PWN formed by \object{PSR~B0833$-$45}, the Vela pulsar (spin-down power of \(\dot{E}\simeq7\times10^{36}\rm\,erg\,s^{-1}\) and characteristic age of \(\tau=1.1\times10^4\rm\,yr\), \citealt{manchester2005}). This association places the PWN at a distance of \(287^{+19}_{-17}\rm\,pc\) from the Earth \citep[][from parallax measurement of the neutron star]{dodson2003}. ROSAT revealed an  X-ray shell with a diameter of 8\degr associated with the supernova remnant (SNR) connected to the Vela pulsar and enclosing Vela X \citep{aschenbach1995}. ROSAT also revealed a 1\degr elongated structure within Vela X that seemingly emanates from the pulsar and was dubbed the cocoon \citep{markwardt1995}\footnote{The choice of name cocoon was motivated by the interpretation, now superseded, that this structure confines particles injected by a jet from the pulsar}. \hess\ is an array of gamma-ray imaging atmospheric Cherenkov telescopes located at an altitude of 1800~m above sea level in the Khomas highland of Namibia. In 2006 it unveiled the TeV emission associated with the cocoon \citep{hess2006velax}, making Vela~X one of the archetypal TeV PWNe.

To explain multi-wavelength observations of Vela X,  \citet{dejager2008} proposed the existence of two distinct electron populations, one corresponding to the extended radio nebula and another to the X-ray/TeV cocoon,  both interacting with magnetic fields of similar strength of $\sim5\rm\,\upmu G$.
\citet{hinton2011} proposed that the extended radio nebula is filled with a relic electron population, devoid of high-energy particles (\(>10\rm\,GeV\)) owing to energy-dependent escape, while the cocoon is filled with electrons accelerated more recently. This scenario is in agreement with hydrodynamical simulations that suggest that the cocoon was formed in an evolved stage of the system, when the reverse shock from the SNR crushed the PWN \citep{blondin2001,slane2018}. These simple two-zone models cannot, however, reproduce the details of the multi-wavelength morphology revealed by the latest observations, such as the larger-scale TeV emission overlapping the extended radio nebula \citep{hessvelax2012}, or hints of energy-dependent morphology at GeV energies \citep{grondin2013}.

In this work we have taken advantage of archival data from the \suz X-ray Imaging Spectrometer (XIS) and of an extended \hess dataset to improve the constraints on the properties of the highest-energy particles and magnetic fields in the Vela X cocoon. The \suz~XIS \citep{koyama2007} has a large field of view with a diameter of \(18'\) and provides spectral coverage in the range from \(0.2\rm\,keV\) to \(12\rm\,keV\) with low background contamination, which enables us to probe emission from the highest-energy electrons in Vela~X \citep[for previous studies, see, e.g.][]{mattana2011,katsuda2011}. The same electrons are also responsible for the multi-TeV emission observed with \hess indicating a cutoff in the underlying particle spectrum \citep{hess2006velax,hessvelax2012}. 

\begin{figure*}[!t]
     \centering
     \includegraphics[width=1.\textwidth]{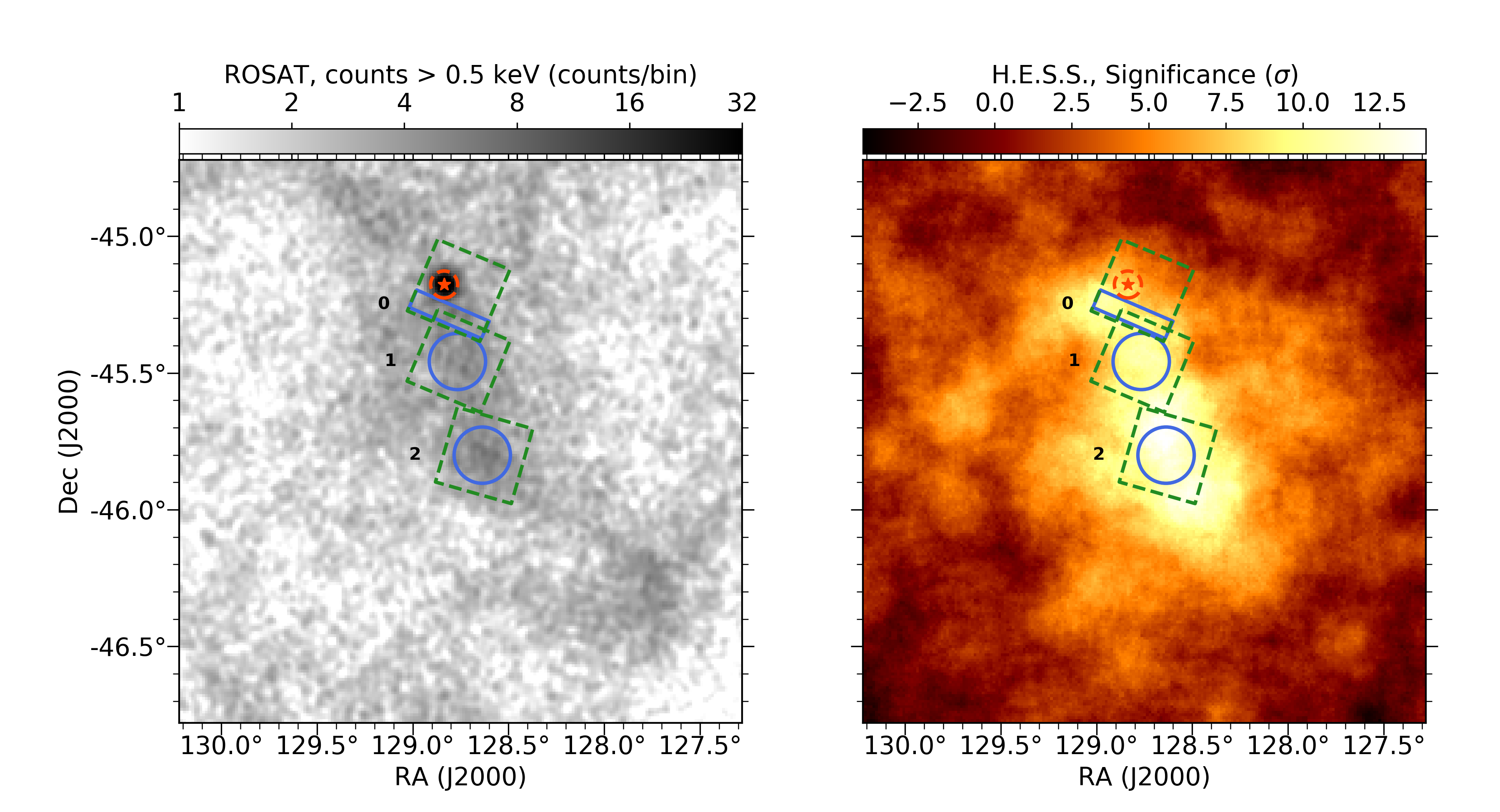}
     \caption{Regions used for spectral analysis (blue box and circles) overlaid onto two maps of the Vela~X region. Left panel: X-ray count map from the \textit{ROSAT} survey at energies $>0.5\rm\,keV$. Right panel: significance map from \hess at energies $>0.6\rm\,TeV$ (see Sect.~\ref{sec:gamma} for details on how the map is derived, the map is oversampled with a correlation radius of $0\fdg2$ for display). The red star indicates the position of the Vela pulsar and its size approximately corresponds to that of the jet-torus structure measured with \textit{Chandra} \citep{manzali2007}. The dashed red circle represents the 95\% containment radius of the \suz point spread function around the pulsar. The dashed green squares indicate the borders of the \suz field of view for the three pointings used in this paper.}
     \label{fig:maps}
\end{figure*}

This energy range is very interesting because the shape of the cutoff encodes information on the particle acceleration and transport mechanism \citep[see][and references therein]{zirakashvili2007,romoli2017}, and also because the highest X-ray energies may reveal a spectral hardening from non-uniform-strength magnetic fields \citep{kelner2013}. Owing to the proximity of Vela~X, and subsequently its large apparent size, and its extremely high flux at TeV energies, we can now extract spatially resolved spectra extending to the highest energies for compact regions in X-rays and gamma~rays, covering different distances from the pulsar wind termination shock, thus overcoming some limitations that affect multi-wavelength studies of other PWNe. 

This paper is structured as follows: in Section~\ref{sec:obs} we describe the observations and the definition of the analysis regions. Then we describe the X-ray and gamma-ray data analysis in Section~\ref{sec:xray} and Section~\ref{sec:gamma}, respectively. In Section~\ref{sec:mwl} the multi-wavelength spectral energy distributions (SEDs) are used to constrain the properties of particles and magnetic fields in the cocoon. Finally, we discuss the results in Section~\ref{sec:discussion} and summarise our conclusions in Section~\ref{sec:summary}.   

\section{Observations and analysis region definition}\label{sec:obs}

We used three archival \suz observations of the Vela X cocoon conducted in 2006. From north (closer to the pulsar) to south, they have observation IDs 501109010 (hereafter Pointing~0), 501107010 (hereafter Pointing~1), 501110010 (hereafter Pointing~2), for an exposure of 60~ks, 61~ks, and 18~ks, respectively.

We define three regions for spectral analysis corresponding to the \suz pointings as illustrated in Figure~\ref{fig:maps}. Later we use these very same regions to derive a SED from the X-ray data and the gamma-ray data.
For Pointings~1 and~2 we used circular spectral extraction regions with a radius of 7.5\arcmin centred in the middle of the \suz~XIS field of view (R.A. = 128\fdg7666, Dec. = $-45\fdg4581$, and R.A. = 128\fdg6368, Dec. = $-45\fdg8007$, respectively). For Pointing~0 we defined a spectral extraction region such that we were able to avoid the region immediately adjacent to the pulsar. The pulsar itself emits X-rays up to a few~keV due to thermal emission from its surface, and to higher energies from magnetospheric particle acceleration and radiation \citep{pavlov2001}. Furthermore, within 1.5\arcmin (\(0.13\rm\,pc\)) from the star, the pulsar wind creates a complex jet-torus structure very bright in X-rays \citep{kargaltsev2003,manzali2007}, not resolved in gamma rays. We excluded from the analysis a circular region centred on the pulsar with a radius of 3.6\arcmin (\(0.3\rm\,pc\)), which corresponds to the 95\% containment radius of the \suz~XIS point spread function (PSF). The final spectral extraction region in Pointing~0 is therefore defined as a rectangle centred at R.A. = 128\fdg81, Dec. = $-45\fdg286$, with major edge of 17.4\arcmin, minor edge of 4.2\arcmin, and tilted by 157\degr with respect to the R.A. axis.  

From 2004 to 2012 \hess was operated as an array of four telescopes with 12~m diameter. In 2012 a fifth telescope with 28~m diameter was added, improving the performance at low energies. We used a dataset spanning from 2004 to 2016, comprising observations taken as part of the survey of the Galactic plane \citep{hgps2018}, from studies of nearby sources such as the Vela pulsar \citep{HESSVelaPSR2018}, Puppis~A \citep{hess2015pupa}, and Vela~Junior \citep{hess2016velajr}, as well as dedicated observations.

We applied the following selection criteria to the dataset:
\begin{itemize}
\item observations for which at least three of the \(12\rm\,m\) telescopes were operational to improve high-energy performance, since we are interested in gamma~rays at energies $>1\rm\,TeV$, produced by the same electrons responsible for the emission measured by \suz in the keV range;
\item observations free from occasional hardware failures (i.e. malfunctioning camera pixels) to closely match the nominal description of the instruments in the Monte~Carlo simulations used to evaluate the array performance;
\item observations with good atmospheric transparency, determined by a cut on the `Cherenkov transparency coefficient' \citep{hahn2014} so that the observing conditions are also well-matched to the description of the atmosphere in the Monte~Carlo simulations.
\end{itemize}
The selection yields 100~h of livetime spent observing within $3\fdg5$ from the centre of Vela X, at R.A. = 128\fdg75, Dec. = $-45\fdg6$ (all coordinates are given in J2000 equinox in this paper).

\section{X-ray analysis}\label{sec:xray}

\begin{figure}[!htbp]
\centering
\includegraphics[width=0.5\textwidth]{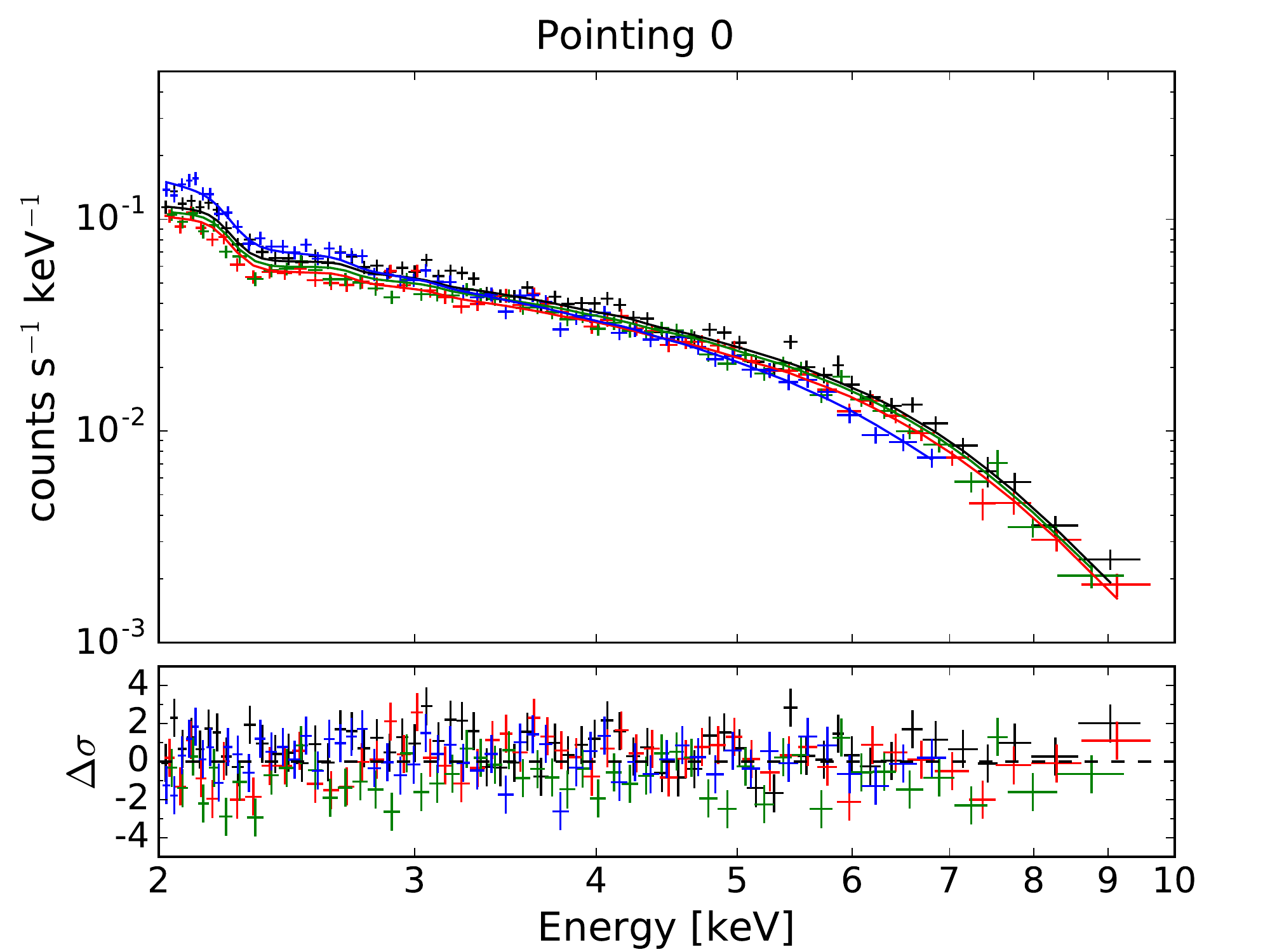}
\includegraphics[width=0.5\textwidth]{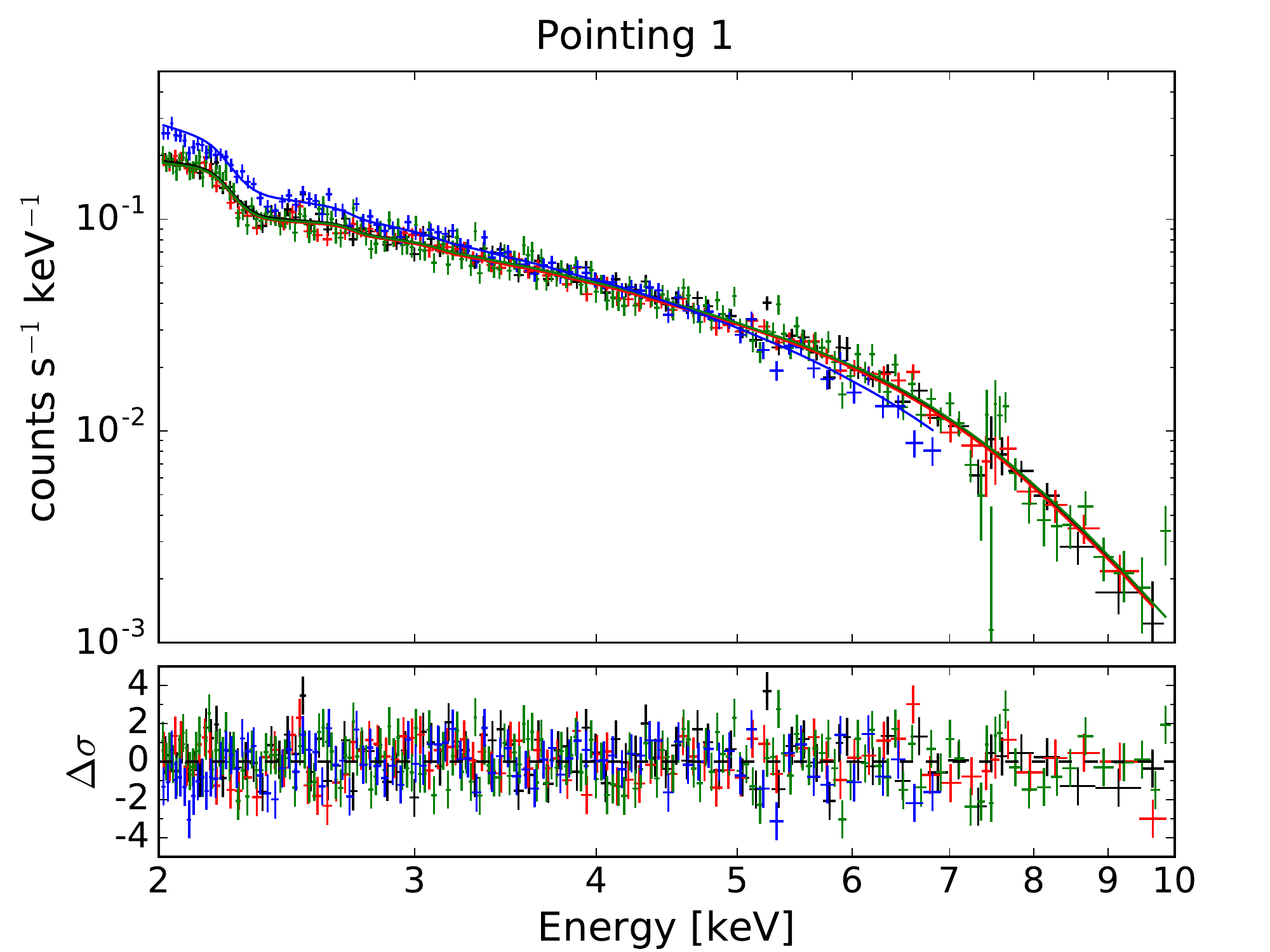}
\includegraphics[width=0.5\textwidth]{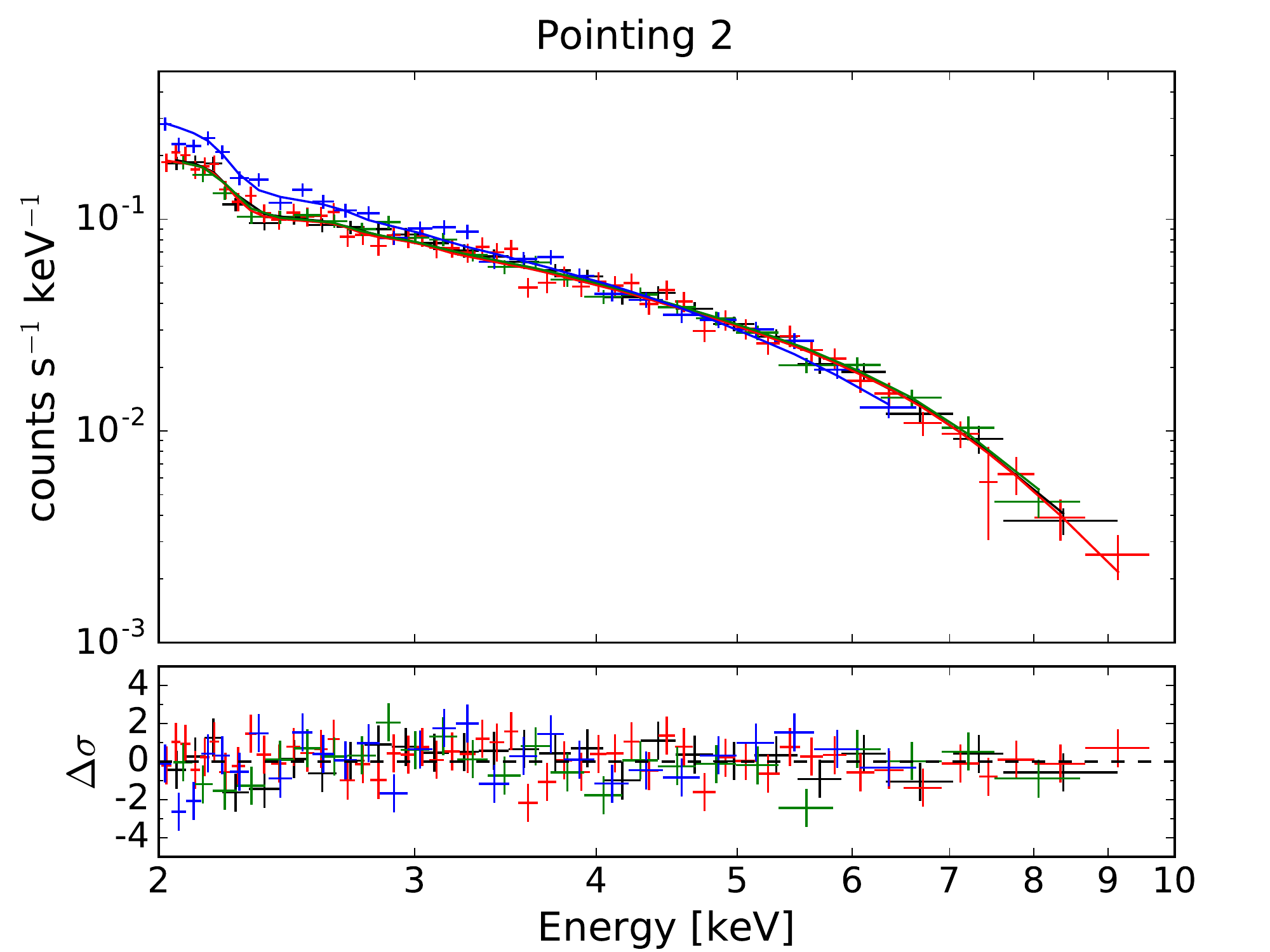}
\caption{\suz X-ray count spectra measured in the regions as illustrated in Figure~\ref{fig:maps}. In each panel, we show spectra obtained with XIS 0 (black), XIS 1 (blue), XIS 2 (red), and XIS 3 (green), superposing the best-fit model as solid lines. The residuals with respect to the best-fit model are shown in the lower panel. Error bars on the flux are 1$\sigma$ statistical uncertainties.}
\label{fig:xray_spectra}
\end{figure}

We used the \texttt{HEASoft} package\footnote{\url{https://heasarc.gsfc.nasa.gov/docs/software/heasoft/}.} version~6.19 for data reduction and analysis.
The data are processed through the \suz standard pipeline in which standard event screening criteria are applied to data obtained with all the XIS charge-coupled devices  (CCDs, XIS 0--3).
We then extract X-ray spectra from the analysis regions in Pointings~0, 1, and 2 (see Figure~\ref{fig:maps}). The instrument response is calculated using the \texttt{xisrmfgen} and \texttt{xissimarfgen} tools.
The instrument response is generated assuming uniform emission that extends to a circular region with a radius of $20'$, which is sufficiently larger than the field of view and a standard setting of \texttt{xissimarfgen} for data analysis of diffuse emission.
We evaluated instrumental background based on night Earth's observations by using \texttt{xisnxbgen}.
We used only photons with energies above \(2.25\rm\,keV\) to avoid thermal emissions from the Vela SNR, and set an upper energy limit at \(10\rm\,keV\) for front-illuminated CCDs (XIS 0, 2, 3) and to \(7\rm\,keV\) for the back-illuminated CCD (XIS 1) to minimise effects from particle-induced instrumental background.

Each X-ray spectrum is well described by a single power law modified by a fixed Galactic interstellar absorption $N_\mathrm{H}=2.59\times 10^{20}\rm\,cm^{-2}$ \citep{manzali2007} plus underlying cosmic X-ray background (CXB), parametrised as in \citet{miyaji1998}, that is, a power law with a photon index of 1.42 and a normalisation of $10.0\rm\,s^{-1}\,cm^{-2}\,sr^{-1}\,keV^{-1}$ at \(1\rm\,keV\). We note that, owing to the proximity of the system and the high-energy threshold, the effect of interstellar absorption on the Vela~X parameters is negligible.
A potential contamination from the Galactic ridge X-ray emission is negligible in this region as Vela~X is sufficiently far from the Galactic centre.
The spectra and best-fit parameters obtained from the fits are shown in Figure~\ref{fig:xray_spectra} and Table~\ref{table:xray_parameters}, in which all power-law fluxes are calculated over an energy range of \(2\)--\(10\rm\,keV\).

The fit results show no significant differences between Pointings~1 and 2 in the flux level or spectral slope ($\sim2.2$--$2.3$), consistent with \textit{XMM-Newton} results in \citet{slane2018}, who find that the X-ray spectrum softens at angular distances exceeding \(60'\) from the pulsar, that is, beyond the bright TeV cocoon. Conversely, in Pointing~0 we find a significantly harder spectrum (photon index $1.92\pm0.014$) and the emission is a factor of greater than two brighter than in the other two regions (the flux within the extraction region is similar, but the solid angle subtended is 41\% of that in regions~1 and~2). 
This is consistent with previous studies of X-ray emission in the region with \textit{XMM-Newton} and \textit{BeppoSAX} \citep{mangano2005}.

Based on the spectral analysis, we derived X-ray SED points in ten energy bins between \(2.25\rm\,keV\) and \(9.5\rm\,keV\).
We included in the subsequent multi-wavelength analysis a 10\% systematic uncertainty on flux measurements \citep{sekiya2016}.
Furthermore, we include additional uncertainties to the SED points of Pointing~0 due to the leakage into the spectral extraction region from emission close to the pulsar. We estimated this uncertainty contribution as the spill-over due to the XIS PSF of the flux from a point source at the position of the pulsar that accounts for the total flux measured in the circular region with radius of 3.6\arcmin centred on the neutron star. It amounts to a fraction of the flux that varies from 35\% to 50\% going from low to high energies and it constitutes the dominant source of uncertainty for Pointing 0.

\begin{table}[!htbp]
\caption{Parameters of the spectral fits to X-ray data}
\label{table:xray_parameters}
\centering
\begin{tabular}{ccc}
\hline\hline
Region & Photon index & Flux\tablefootmark{a} ($10^{-7}\rm\,erg\,s^{-1}\,cm^{-2}\,sr^{-1}$)\\
%\tnote{a} \\
\hline
0 & $1.920\pm0.014$ & $13.86\pm0.15$ \\
1 & $2.251\pm0.013$ & $5.72 \pm0.05$ \\
2 & $2.31\pm0.02 $&   $5.66 \pm0.10$ \\
\hline
\end{tabular}
\tablefoot{For fixed $N_\mathrm{H}$ and CXB parameters, see text for details. The effect of interstellar absorption is negligible. All errors are reported as $1\sigma$ confidence intervals.\\
\tablefoottext{a}{The flux within the extraction region is reported for the energy range \(2\rm\,keV\) to \(10\rm\,keV\).}}
\end{table}%

\section{Gamma-ray analysis}\label{sec:gamma}

   \begin{table*}[!bthp]
      \caption{Best-fit parameters for the spectral models of the gamma-ray spectrum of the Vela X cocoon in the three regions shown in Figure~\ref{fig:maps}. Errors correspond to $1\sigma$ uncertainties. The functional forms used to model the spectra are given in Equations~\ref{eq:specPL} and~\ref{eq:specEXP}. TS is the test statistic for the improvement of the fit when the ECPL model is used, see text for definition.}
         \label{tab:fit-param} 
         \centering
         \begin{tabular}{lccccc}
            \hline\hline
            Pointing            & Model & Flux $> 1\rm\,TeV$                            & $\Gamma$                & $E_\mathrm{cut}$      & TS\\
                                &               & ($10^{-8}\rm\,cm^{-2}\,s^{-1}\,sr^{-1}$)      &                               & (TeV) &\\
            \hline
            0                   & PL            & $3.6 \pm 0.4_\mathrm{stat} \pm 0.7_\mathrm{sys}$   & $1.75 \pm 0.10_\mathrm{stat} \pm 0.14_\mathrm{sys}$   & -- & --\\
                                & ECPL  & $3.5 \pm 0.5_\mathrm{stat} \pm 0.7_\mathrm{sys}$       & $1.2 \pm 0.3_\mathrm{stat} \pm 0.1_\mathrm{sys}$      &       $16 \pm 7_\mathrm{stat} \pm 6_\mathrm{sys}$ & 6.4\\
            \hline
            1                   & PL    & $4.1 \pm 0.2_\mathrm{stat} \pm 1.0_\mathrm{sys}$       & $1.92 \pm 0.06_\mathrm{stat} \pm 0.10_\mathrm{sys}$   &       --      & --\\
                                & ECPL  & $4.1 \pm 0.3_\mathrm{stat} \pm 1.5_\mathrm{sys}$       & $1.47 \pm 0.16_\mathrm{stat} \pm 0.16_\mathrm{sys}$   &       $14 \pm 5_\mathrm{stat} \pm 6_\mathrm{sys}$ & 13.4\\
            \hline
            2                   & PL    & $4.6 \pm 0.3_\mathrm{stat} \pm 0.9_\mathrm{sys}$       & $1.84 \pm 0.06_\mathrm{stat} \pm 0.11_\mathrm{sys}$ & --    & -- \\
                                & ECPL  & $4.6 \pm 0.3_\mathrm{stat} \pm 1.4_\mathrm{sys}$       & $1.27 \pm 0.14_\mathrm{stat} \pm 0.10_\mathrm{sys}$   &         $12 \pm 3_\mathrm{stat} \pm 6_\mathrm{sys}$     & 25.7\\
            \hline
         \end{tabular}
   \end{table*}
   
\begin{figure*}[!bthp]
     \centering
     \includegraphics[width=1.\textwidth]{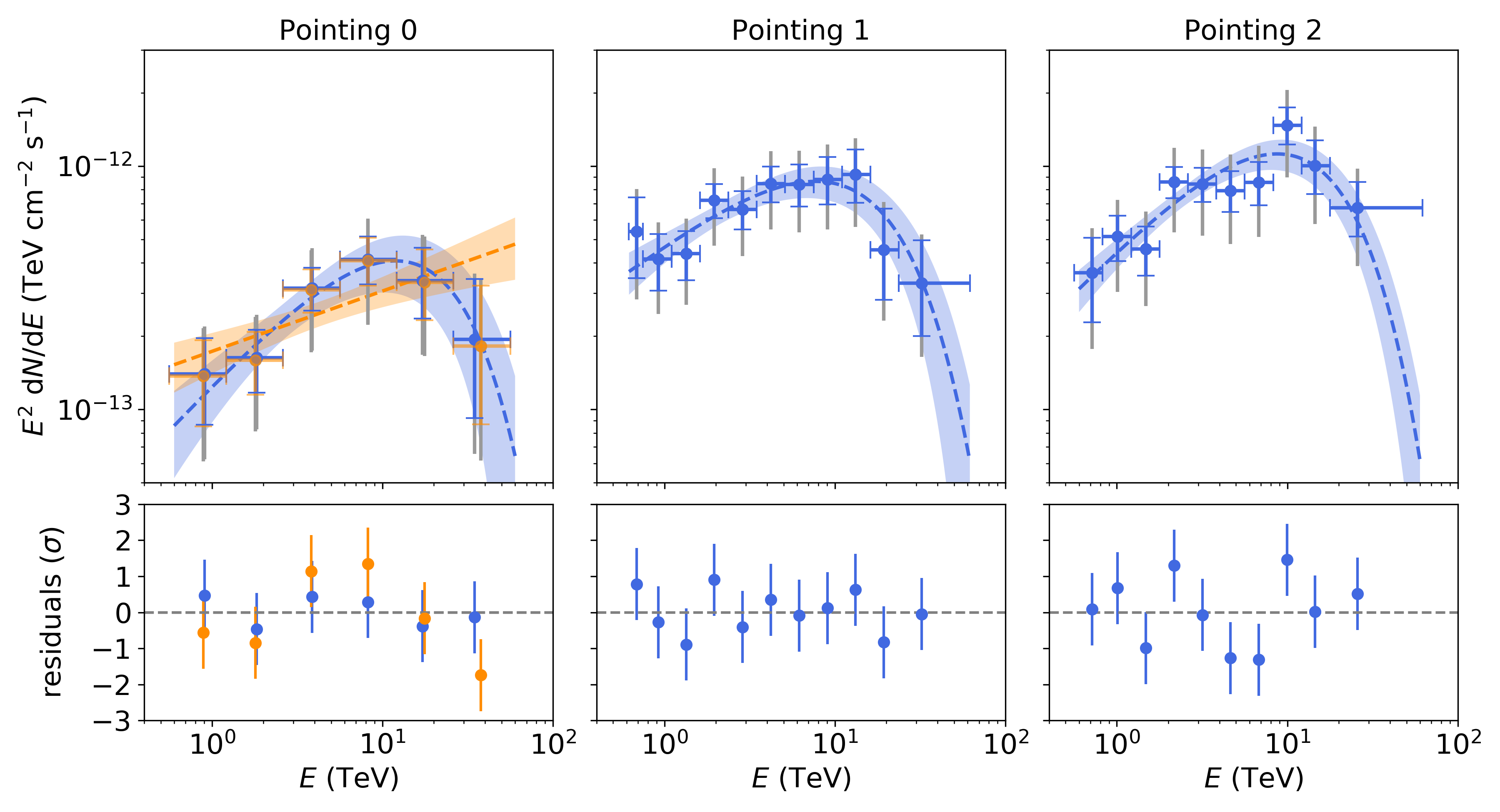}
     \caption{SEDs of gamma-ray emission measured by \hess for the three extraction regions shown in Figure~\ref{fig:maps}. In the top panel the lines show the best-fit spectral models, and the bands display the statistical uncertainties based on quadratic error propagation from the inversion of the likelihood Hessian matrix. Points show the binned SEDs with their statistical uncertainties (coloured error bars with end caps) and the sum in quadrature of statistical and systematic uncertainties (grey error bars without end caps). The bottom panel shows the deviation of the SED points from the best-fit function as number of statistical $\sigma$. For Pointing~0, orange corresponds to the PL fit, and blue to the ECPL fit.}
     \label{fig:spec_hess}
\end{figure*}

Since we are mainly interested in the energy range $> 1\rm\,TeV$, to minimise systematics that may be induced by combining different instrumental configurations we use only data from the four 12~m telescopes for the whole time period (the contribution from the 28-m telescope to the effective area in the energy range of interest is unimportant). The gamma-ray energy and direction reconstruction, and separation from the background of cosmic-ray hadrons are based on a multivariate technique applied to parameters of the Cherenkov images of the atmospheric showers \citep{ohm2009}. Given the large extension of Vela~X and low surface brightness, we used tight selection criteria to make the residual contamination from hadrons misclassified as gamma rays as low as possible, in other words, we applied the `hard cuts' from \citet{ohm2009}.  Additionally, to ensure uniform energy thresholds between different observing periods, we included in the subsequent analysis only candidate gamma-ray events with reconstructed energy $>0.6\rm\,TeV$, thus reducing systematics in spectral reconstruction at the lower end of the energy range. 

All the results in the paper were cross-checked using a second independent software and analysis chain for calibration, event reconstruction, and selection based on an air-shower `model template' approach \citep{denaurois2009}. Simulated image templates are fitted to the measured image, in order to obtain the physical properties of the shower progenitor gamma ray. The goodness of fit is used as a discriminant variable between gamma-ray and background events. {Also in this case we used `hard cuts'} as defined in \cite{aharonian2006crab}. Results are given for the main analysis, and the alternative analysis is used to derive systematic uncertainties when noted.

The right-hand panel of Figure~\ref{fig:maps} shows the gamma-ray detection significance for the whole region of the Vela~X cocoon. The map was derived using the formalism by \citet{lima1983} after evaluating the residual hadronic background using the ring method \citep[e.g.][]{berge2007}. In the latter, the residual background was estimated from the data within the same observation, excluding regions of the field of view in which significant gamma-ray emission is potentially present. We specifically excluded from the background estimation regions that had gamma-ray emission detected at significance $>5\sigma$ in the \hess Galactic plane survey \citep[][3.2.2]{hgps2018}, or for which the brightness temperature at 44~GHz measured with \textit{Planck} \citep{planck2014} is $> 1.5$~mK, which indicates the presence of relativistic electron populations that may radiate in gamma rays\footnote{The second criterion increases the area of the excluded region by $<5\%$.}. Figure~\ref{fig:maps} shows that there is very significant gamma-ray emission at energies $>0.6\rm\,TeV$ in all three analysis regions.

We evaluated the residual background for spectral estimation applying the reflected-region method \citep[e.g.][]{berge2007}. The requirement to have at least two reflected regions for background estimation outside the exclusion region (the same described above for the derivation of the significance map) reduces the livetime to 70~h, 75~h, and 80~h for Pointings~0, 1, and 2, respectively. Taking into account the background count spectra thus estimated, we fitted analytic functions to the count spectra in the spectral extraction regions using a maximum likelihood method based on Poisson statistics for both counts and background counts. In Pointing 0 counting statistics above a few tens of TeV are low, with no events recorded at reconstructed energies $>60\rm\,TeV$, except for one single event at a reconstructed energy of $\sim90\rm\,TeV$ in the spectral extraction region (no events with reconstructed energies $>60\rm\,TeV$ are found in the background regions). We have verified that all the results presented in the paper are not affected, within statistical uncertainties, by including the $\sim90\rm\,TeV$ event in the spectral analysis or not. We report in the following the results obtained by selecting only events with reconstructed energy $<60\rm\,TeV$ in Pointing~0. 

We considered two analytical functions to represent the source photon spectrum. The first is a simple power law (PL), for which the number of photons, $N$, as a function of energy, $E$, varies as
\begin{equation}\label{eq:specPL}
\frac{\mathrm{d}N}{\mathrm{d}E} = A \left(\frac{E}{E_0}\right)^{-\Gamma},
\end{equation}
where $A$ is the differential flux that normalises the spectrum at $E_0$, and $\Gamma$ is the spectral index of the distribution.
Alternatively we considered an exponentially-cutoff power law (ECPL)
\begin{equation}\label{eq:specEXP}
\frac{\mathrm{d}N}{\mathrm{d}E} = A \left(\frac{E}{E_0}\right)^{-\Gamma} \exp\left(-\frac{E}{E_\mathrm{cut}}\right),
\end{equation}
with the characteristic energy of the cutoff, $E_\mathrm{cut}$, as additional parameter. We used Eqs.~(\ref{eq:specPL})~and~(\ref{eq:specEXP}) simply to describe the shape of the gamma-ray spectrum. To infer more detailed information about the physical processes behind the gamma-ray emission in Section~\ref{sec:mwl} we determine the energy distribution of high-energy electrons in the system by fitting their non-thermal radiation to \suz and \hess data.

Let $\mathcal{L}_\mathrm{PL}$ and $\mathcal{L}_\mathrm{ECPL}$ be the maximum likelihood values for the PL and ECPL models, respectively, given the source region and background regions counts. The test statistic $\mathrm{TS}=2\log(\mathcal{L}_\mathrm{ECPL}/\mathcal{L}_\mathrm{PL})$ is used to assess whether the additional degree of freedom associated to the cutoff significantly improves the fit. Although the formal criteria to apply the likelihood ratio test  \citep[e.g.][]{protassov2002} are not applicable\footnote{Eq.~\ref{eq:specEXP} reduces to Eq.~\ref{eq:specPL} in the limit $E_\mathrm{cut} \rightarrow +\infty$, therefore the null hypothesis (PL) lies on the border of the parameter space for the test hypothesis (ECPL).}, a larger TS value can still be taken to represent a significant improvement in the fit when using the ECPL model. For Pointings~1 and~2 we obtain TS equal to 13.4 and 25.7, respectively, therefore we selected the ECPL as best-fit model. For Pointing 0 TS is 6.4 formally corresponding to a significance of the cutoff at $2.5\sigma$ statistical level.  The presence of a cutoff in Pointing~0 is not significant. However, we note that the value of the cutoff energy, $16 \pm 7\rm\,TeV$, is consistent with those obtained for the other two Pointings.

The resulting spectral parameters and gamma-ray fluxes for the best-fit functions are reported in Table~\ref{tab:fit-param}.
Systematic uncertainties on the fit parameters are evaluated combining the differences between results from the two different calibration, reconstruction, and event selection schemes used in this work with the other sources of systematic uncertainties as evaluated in \citet{aharonian2006crab}, namely 20\% on flux, 0.1 on the spectral index, and 30\% on the cutoff energy. These include, for the flux, uncertainties due to the Monte~Carlo models of the atmosphere and particle interactions used to evaluate the instrument performance, effect of missing camera pixels, uncertainty in the livetime estimate, and, for all parameters, the effect of the choice of event selection criteria and background estimation method, and systematic fluctuations observed either within single runs or on a run-by-run basis. The results from the two different calibration, reconstruction, and event selection schemes are consistent with each other, with the exception of the cutoff energy in Pointing~2, which differs by $\sim3\sigma$. This deviation is taken into account in the systematic uncertainties. The spectral parameters are consistent with previous analyses of \hess data covering larger regions of the Vela~X cocoon \citep{hess2006velax,hessvelax2012}.
Additional information on the spectral fitting is provided in Appendix~\ref{app:hessfit}.
   
The results from the spectral fitting are used to derive a binned SED for each pointing. The spectra are rebinned such that for each point the significance of gamma-ray signal detection is $>2\sigma$, and we require that each bin has at least two excess\footnote{The excess is defined as the number of events in the spectral extraction region minus the background estimate derived from the events in the background regions.} counts. The binned SEDs are shown along with the full-range spectral functions in Figure~\ref{fig:spec_hess}. For Pointings~1 and 2 we consider only the ECPL hypothesis, which is strongly supported by the data, while for Pointing~0 we show both the results based on the PL and ECPL hypotheses. Systematic uncertainties on the SED points are shown as the sum in quadrature of the systematic error on flux from the sources discussed above, amounting to 20\%, with the average difference between the points obtained from the two analyses. The average is performed over energy independently for every pointing, and it amounts to 35\%, 24\%, and 29\% for Pointing 0, 1, and 2, respectively. Figure~\ref{fig:spec_hess} shows that within statistical uncertainties the binned SEDs agree with the best-fit function determined from the whole energy range. For Pointing 0 we note that the SEDs derived based on the PL and ECPL hypotheses are consistent within the level of the statistical uncertainties. Since the cutoff model is not statistically favoured, for this pointing in Section~\ref{sec:mwl} we have adopted the SED derived from the PL fit. We have verified that all the results are consistent with those obtained from the ECPL SED within the quoted uncertainties.

\section{Radiative modelling of the multi-wavelength spectral energy distribution}\label{sec:mwl}

To study the obtained X- and gamma-ray spectra in a consistent way, we computed the synchrotron and IC emission from a population of high energy electrons. The energy distribution of particles is assumed to be a power-law with sub or super-exponential cutoff:
\begin{equation}\label{eq:espec}
\frac{\mathrm{d}N_e}{\mathrm{d}E_e} = A \left(\frac{E_e}{E_0}\right)^{-\alpha} \exp{\left[-\left(\frac{E_e}{E_\mathrm{cut}}\right)^\beta\right]},
\end{equation}
where $N_e$ is the electron number, $E_e$ is the electron energy, $A$ is the normalisation factor for the distribution, $\alpha$ is the spectral index of a power-law distribution with reference energy $E_0$, $E_\mathrm{cut}$ is the cutoff energy, and $\beta$ is the cutoff index. 

Leptons produce X-ray emission through synchrotron radiation in a magnetic field, which is assumed to have a random
orientation but uniform strength $B$.  The gamma-ray emission is generated through IC scattering on CMB and FIR photon
fields. For the latter we use a recent model by \citet{popescu2017} at the position of Vela~X. The relevance of FIR
radiation, which was often overlooked in past studies, and the possible uncertainties imposed by the model for the FIR
field as seed for the IC scattering in Vela~X are discussed in Appendix~\ref{app:irfield} \cite[for a general discussion of the contribution of different photon fields to IC scattering in extended gamma-ray sources see also][]{1997MNRAS.291..162A}.  

The computation of the SED models and subsequent fit to the multi-wavelength SEDs are performed using the \texttt{naima}
Python package \citep{zabalza2015}. Specifically, synchrotron emission is computed based on the formalism in
\cite{aharonian2010}, and IC emission on the formalism in \cite{khangulyan2014,aharonian1981}.

The \texttt{naima} package enables us to derive the best-fit values and posterior probability distributions of the model parameters given the SED points from the $\chi^2$, calculated assuming that the SED point uncertainties are Gaussian and uncorrelated. Owing to the presence of systematic uncertainties, that, to this end, we combine in quadrature with statistical ones, and to the instruments' energy dispersion, this is only an approximate assumption, that should be overcome in future works through multi-mission full-forward analyses \citep[e.g.][]{vianello2015}. The model parameters are scanned using the Markov~chain Monte~Carlo (MCMC) method, as implemented in the \texttt{emcee} software package \citep{foreman2013}. For all model parameters we assume a flat prior probability distribution, within physical constraints on the parameters values (e.g. particle densities are positive). We scanned the cutoff energy $E_\mathrm{cut}$ in logarithmic space, so that the fit parameter is actually $\log_{10}(E_\mathrm{cut}/1\,\mathrm{TeV})$.

\begin{figure}[!tbhp]
\centering
\includegraphics[width=0.5\textwidth]{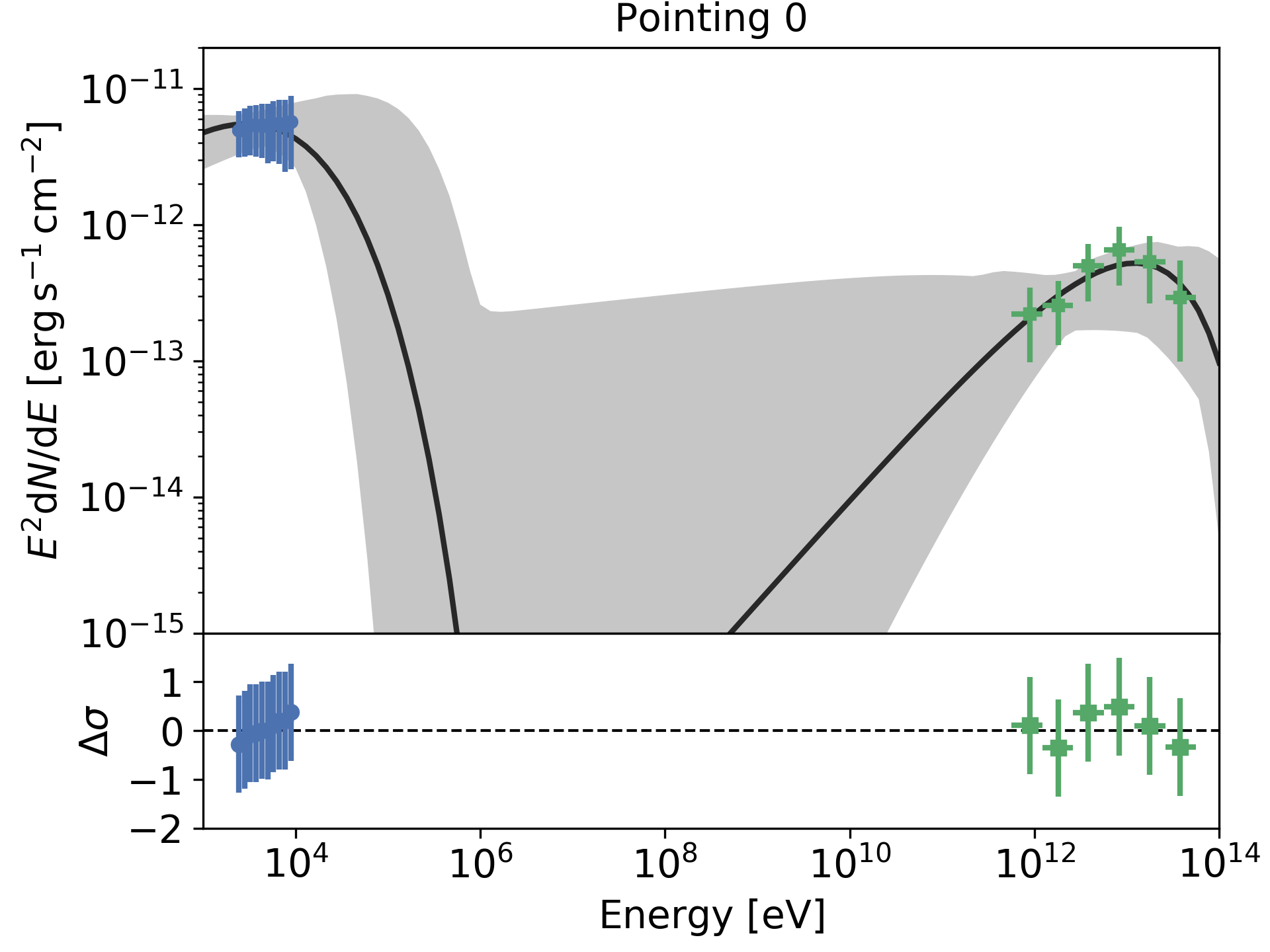}
\includegraphics[width=0.5\textwidth]{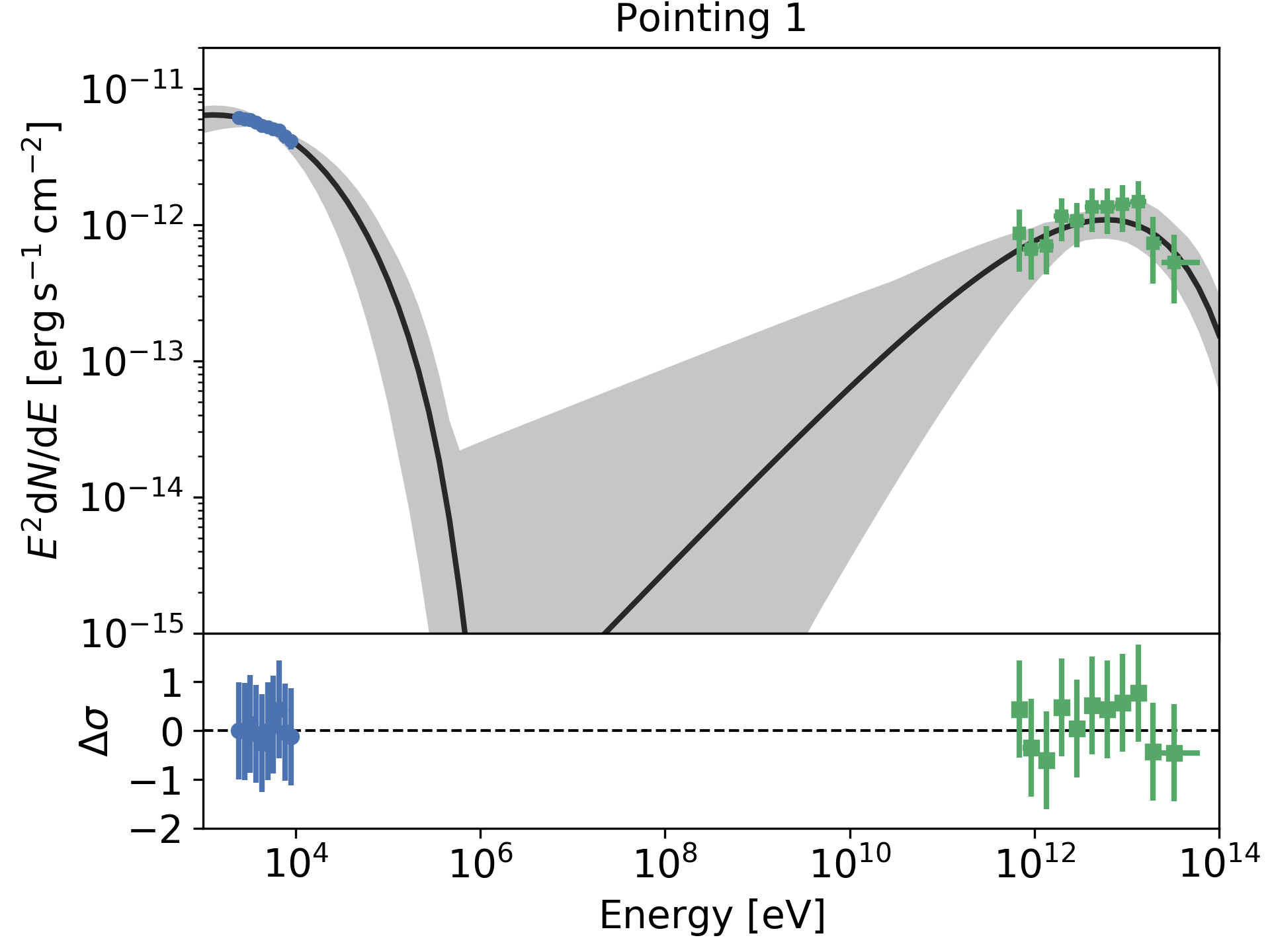}
\includegraphics[width=0.5\textwidth]{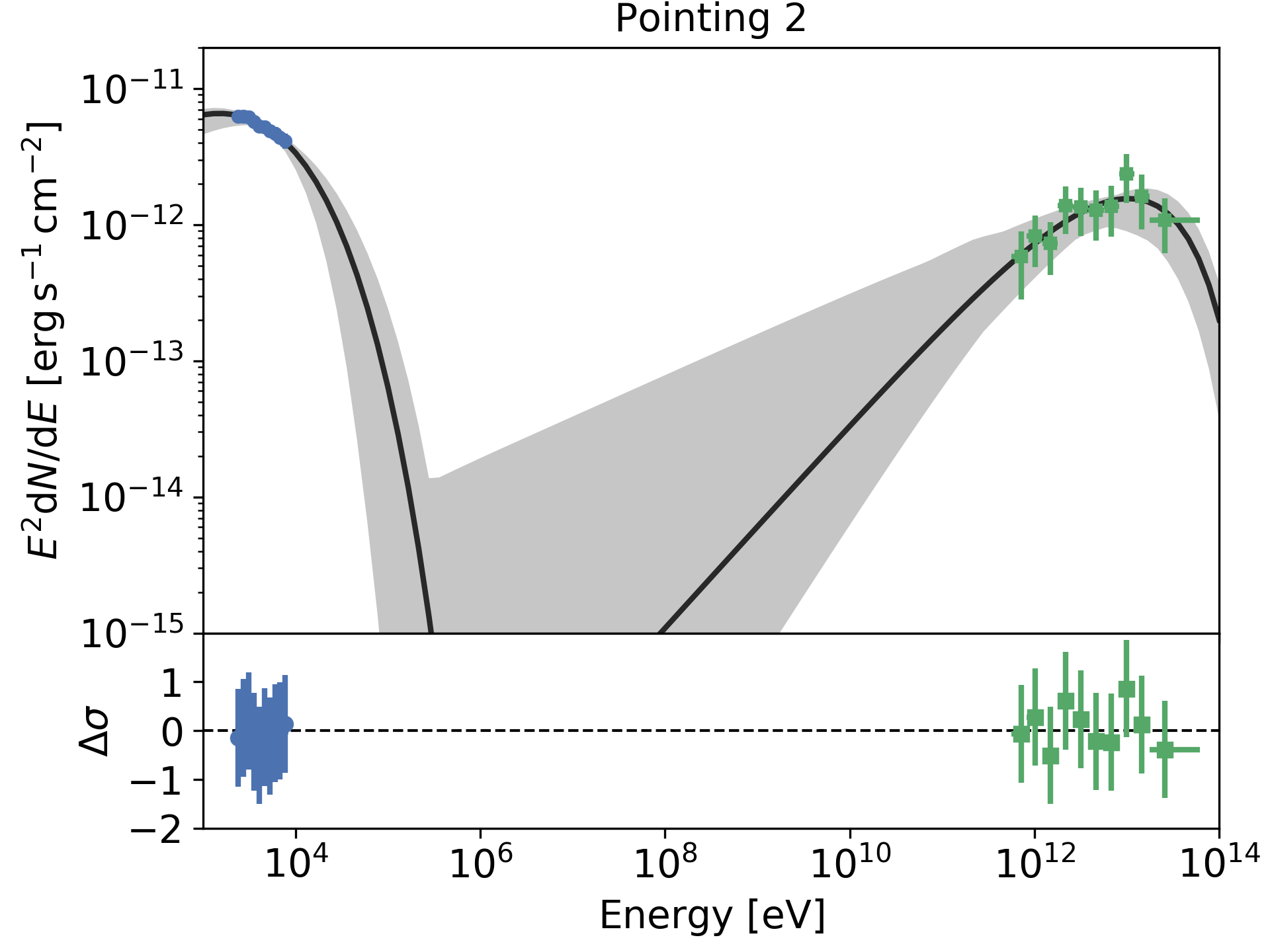}
\caption{Multi-wavelength spectral-energy distributions (SEDs) from the three regions studied in the Vela~X cocoon. Blue points are derived from X-ray measurements with the \suz~XIS (Section~\ref{sec:xray}), and green points from the gamma-ray measurements with \hess (Section~\ref{sec:gamma}). Error bars combine statistical and systematic uncertainties as described in text. Top, middle, and bottom panels correspond to the extraction regions 0, 1, and 2, respectively. In the upper sub-plot of each panel, radiative models  are overlaid to the SED points. In the lower subplot, the residuals for the best fit parameters are shown. The models are based on a single population of leptons producing synchrotron emission in a uniform-strength magnetic field and inverse-Compton emission on the cosmic microwave background and diffuse infrared radiation. The black line shows the best-fit (minimum $\chi^2$) model, while the shaded bands represent the 95\% confidence level bands from the Markov~chain Monte~Carlo scan.}
\label{fig:mwlsed}
\end{figure}

The fit results are shown in Figure~\ref{fig:mwlsed} and Table~\ref{tab:fit-param-mwl}. Figure~\ref{fig:mwlsed} shows the best-fit model SED and model uncertainties compared to the measured SEDs. Table~\ref{tab:fit-param-mwl} gives median and upper and lower uncertainties on the parameter values. In addition, Appendix~\ref{app:corner} contains the probability density distributions of the model parameters. 
   
Figure~\ref{fig:mwlsed} shows that leptonic models can reproduce the X-ray and gamma-ray SEDs for plausible model parameter values (see Table \ref{tab:fit-param-mwl}). The inferred properties of the electron spectra and strength of the magnetic field are consistent within the uncertainties over distances from \(0.3\rm\,pc\) up to \(4\rm\,pc\) from the pulsar wind termination shock. As expected for IC radiation in the Thomson regime, the values of $\alpha$ in Table~\ref{tab:fit-param-mwl} and those of $\Gamma$ in Table~\ref{tab:fit-param} are consistent with the equation $\Gamma=(1+\alpha)/2$. The cutoff exponent $\beta$ is only weakly constrained by the data. Furthermore, the constraints on the cutoff for Pointing~0 are weaker than for the other regions studied, consistent with the harder X-ray spectrum (Section~\ref{sec:xray}) and the lack of a significant cutoff detection in gamma rays (Section~\ref{sec:gamma}). For Pointing~0 the weaker constraints also stem from the larger systematic uncertainties on the X-ray SED, increasing with increasing energy due to the spill-over from the region immediately around the pulsar.

Although we used the X- and gamma-ray spectra extracted from the compact regions, the emission is not necessarily produced in completely homogeneous zones. The spectrum extraction regions span distances of \(\sim1\rm\,pc\), which significantly exceeds the gyro radius of TeV particles. Thus, it cannot be excluded that some MHD inhomogeneity affects the magnetic field strength across the production region. It is important to note that the magnetic field strength can also vary along the line of sight, and that future improvement of the X-ray and gamma-ray instruments' sensitivity will not  remove this uncertainty completely.  

If the magnetic field varies in the production area, the synchrotron spectra can be deformed significantly \citep{kelner2013}. The key parameter that determines the radiation regime in an inhomogeneous magnetic field is  the ratio of the magnetic field fluctuation length, $\lambda_B$, to the characteristic photon formation length, $\lambda_\mathrm{ph}=m_e c^2 / e B$ \citep[with $m_e$ electron mass, $e$ electron charge, $c$ speed of light, and $B$ strength of the magnetic field, see][]{1979rpa..book.....R}. Landau damping imposes a rather fundamental constraint on the minimum length of inhomogeneities:    
\begin{equation}
{\lambda_B \over \lambda_\mathrm{ph}} > {\sqrt{{\epsilon}\over 4\pi e^2 n_e}{eB\over m_ec^2} }={\epsilon \over m_ec^2}{\sqrt{B^2\over 4\pi \epsilon n_e}} ,
\end{equation}
where $n_e$ and $\epsilon$ are electron number density and mean energy per electron, respectively. Since \(\epsilon / (m_ec^2)\sim\Gamma_\textsc{PW}\gg 1\) (\(\Gamma_\textsc{PW}\) is the bulk Lorentz factor of the pulsar wind) and \({B^2/ (4\pi \epsilon n_e})\sim 1\) (for energy equipartition between magnetic fields and particles expected in PWNe, \citealt{KenCor84}), one obtains that $\lambda_B / \lambda_\mathrm{ph} \gg 1$.
Thus, in PWNe the impact of magnetic field inhomogeneities on the X-ray spectra can be modelled as a superposition of synchrotron spectra produced in magnetic fields of different strength \citep{kelner2013}. 

Therefore, we gauged magnetic turbulence in Vela~X against the multi-wavelength SEDs by assuming that the probability density function (PDF) corresponding to a magnetic field strength $B$ is:
\begin{equation}\label{eq:turbB}
\mathrm{PDF}(B) = (1-a) \, \delta(B-B_0) + a \, C \, B^{-\zeta} \mathcal{H}(B-B_\mathrm{min}) \, \mathcal{H}(B_\mathrm{max} - B).
\end{equation} 
The first term of the summation represents a magnetic field with a fixed strength, i.e. $B_0$, with $\delta$ being the Dirac distribution. The second term in the summation represents a magnetic field with power-law distribution of index $\zeta$ between minimum and maximum strengths  $B_\mathrm{min}$ and $B_\mathrm{max}$, respectively, with $\mathcal{H}$ being the Heaviside step distribution. The parameter $a$ represents the mixing between the constant and power-law magnetic field components.

For illustration we take $B_\mathrm{max} = 100 \times B_0$, that is, sufficiently exceeding the mean value to modify the synchrotron spectrum, and $\zeta = 3/2$. Although this is a typical power-law index of interstellar MHD turbulence (Kraichnan-type turbulence), we note that the parameter $\zeta$ in our formulation has a different physical meaning, related to the distribution of the magnetic-field strength. The effect of choosing a different power-law index, for example $\zeta = 5/3$,
has an impact on the results that is negligible compared to other sources of uncertainty.

The parameters $C$ and $B_\mathrm{min}$ in Eq.~(\ref{eq:turbB}) are set respectively by requiring that the integral of the PDF is~1, and that the root mean square expectation value for the power-law magnetic field (and, therefore, the total magnetic field as well) is $B_0$ (i.e. $B_\mathrm{RMS}=B_0$). The only free parameters are then the mixing $a$ and the root mean square magnetic field expectation value, $B_0$, which can be fitted to the multi-wavelength SEDs.

   \begin{table*}[!bthp]
      \caption{Best-fit parameters for the radiative model of the X-ray and gamma-ray spectrum of the Vela X cocoon in the three regions shown in Figure~\ref{fig:maps}.}
         \label{tab:fit-param-mwl} 
         \centering
         \begin{tabular}{lcccccc}
            \hline\hline
            Pointing            &  $W_e$ ($> 1\rm\,TeV$)\tablefootmark{a}                               & $\alpha$\tablefootmark{b}               & $E_\mathrm{cut}$\tablefootmark{b} & $\beta$\tablefootmark{b} & $B$\tablefootmark{c} & BIC\tablefootmark{d} \\
                                & ($10^{44}$ erg)       &                               & (PeV) & & ($\upmu$G)\\
             \hline
            0                   & $0.7_{-0.4}^{+1.1}$ & $2.2_{-0.6}^{+0.3}$ & $0.3_{-0.2}^{+4.0}$ & $2.0_{-1.4}^{+2.6}$ & $8.6_{-1.7}^{+2.8}$ & 14.8\\
            1                   & $1.7_{-0.6}^{+0.8}$ & $1.8\pm0.5$ & $0.05_{-0.03}^{+0.07}$ & $0.9_{-0.2}^{+0.5}$ & $6.7_{-0.7}^{+0.9}$ & 18.2\\
            2                   & $1.6_{-0.5}^{+0.7}$ & $1.9_{-0.6}^{+0.3}$ & $0.11_{-0.06}^{+0.03}$ & $2.0_{-0.8}^{+2.3}$ & $5.4_{-0.6}^{+0.8}$  & 17.3 \\
            \hline
         \end{tabular}
         \tablefoot{Median values from the MCMC scan, with lower and upper uncertainties based on the 16th and 84th percentiles of the posterior distribution.\\
\tablefoottext{a}{Total electron energy for particle energies $>1\rm\,TeV$. We note that the solid angle subtended by region~0 is 41\% of that in regions~1 and~2.}\\
\tablefoottext{b}{Parameters of the electron spectrum as defined in Equation~\ref{eq:espec}.}\\
\tablefoottext{c}{Strength of the magnetic field.}\\
\tablefoottext{d}{Bayesian information criterion, i.e. $k \cdot \ln n + \chi^2$, where $k$ is the number of parameters estimated from the model and $n$ is the number of data points.}}
   \end{table*}

   \begin{table*}[!htbp]
      \caption{Best-fit parameters for the radiative model of the X-ray and gamma-ray spectrum of the Vela X cocoon for the case of turbulent magnetic field in the three regions shown in Figure~\ref{fig:maps}.}
         \label{tab:fit-param-mwl-turb} 
         \centering
         \begin{tabular}{lccccccc}
            \hline\hline
            Pointing            &  $W_e$ ($> 1\rm\,TeV$)\tablefootmark{a}                               & $\alpha$\tablefootmark{b}               & $E_\mathrm{cut}$\tablefootmark{b} & $\beta$\tablefootmark{b} & $B_0$\tablefootmark{c} & $a$\tablefootmark{c} & BIC\tablefootmark{d}\\
                                & ($10^{44}$ erg)       &                               & (PeV) & & ($\upmu$G)\\
            \hline
            0                   & $0.7_{-0.3}^{+0.4}$ & $2.3_{-0.6}^{+0.4}$ & $0.24_{-0.19}^{+0.30}$ & $2.3_{-1.7}^{+2.4}$ & $11\pm3$ & $0.4_{-0.3}^{+0.4}$ & 17.3\\
            1                   & $1.3_{-0.4}^{+0.6}$ & $2.0_{-0.6}^{+0.3}$ & $0.09\pm0.04$ & $2.6_{-1.3}^{+1.9}$ & $7.6\pm0.9$ & $0.39_{-0.14}^{+0.11}$ & 20.5\\
            2                   & $1.4_{-0.3}^{+0.7}$ & $1.7\pm0.4$ & $0.08\pm0.03$ & $2.4_{-1.1}^{+1.5}$ & $6.7_{-1.2}^{+1.0}$ & $0.32_{-0.15}^{+0.14}$ & 19.9 \\
            \hline
         \end{tabular}
         \tablefoot{Median values from the MCMC scan, with lower and upper uncertainties based on the 16th and 84th percentiles of the posterior distribution.\\
\tablefoottext{a}{Total electron energy for particle energies $>1\rm\,TeV$. We note that the solid angle subtended by region~0 is 41\% of that in regions~1 and~2.}\\
\tablefoottext{b}{Parameters of the electron spectrum as defined in Equation~\ref{eq:espec}.}\\
\tablefoottext{c}{Parameters of the magnetic field distribution as defined in Equation~\ref{eq:turbB}.}\\
\tablefoottext{d}{Bayesian information criterion, i.e. $k \cdot \ln n + \chi^2$, where $k$ is the number of parameters estimated from the model and $n$ is the number of data points.}}
\end{table*}

Table~\ref{tab:fit-param-mwl-turb} reports median, upper, and lower uncertainties on the parameter values.
Results for the electron spectra parameters and magnetic field expectation value are generally consistent with those derived for a fixed magnetic field (Table~\ref{tab:fit-param-mwl}).
The mixing parameter $a$ is not constrained by the observations in
Pointing~0, while for the other two regions small contributions from the power-law magnetic field distribution are
favoured: the 99th percentiles of the posterior probability distribution of \(a\) lie below the values 0.57 and 0.56 (for
Pointings~1 and~2, respectively). The Bayesian information criterion (BIC) values in Tables~\ref{tab:fit-param-mwl} and~\ref{tab:fit-param-mwl-turb} are smaller for the fixed magnetic field case. Thus, this study does not allow us to claim a detection of the emission generated in a turbulent magnetic field. This is largely unsurprising, since the magnetic field turbulence predominantly affects the high energy part, in other words, a region above SED peak which is not well covered by current X-ray measurements. This also results in large uncertainties on the cutoff index $\beta$, which is the electron spectrum parameter most coupled with the effect of turbulent magnetic fields. This prevents us from studying in detail the shape of the particle spectrum cutoff, that encodes information about the particle acceleration and transport mechanisms \citep[see][and references therein]{zirakashvili2007,romoli2017}.

\section{Discussion}\label{sec:discussion}

\subsection{Pressure balance in the cocoon}
As shown in Table~\ref{tab:fit-param-mwl}, the electron distribution and magnetic field strength are consistent within the uncertainties between the three regions  considered. For a magnetic field strength of $\simeq 6\,\rm\upmu G$ the energy density is 
\begin{equation}\label{eq:b_energy_density}
w_\textsc{b}={B^2\over 8\pi}\simeq1.4\times10^{-12}\rm\,erg\,cm^{-3}\,.
\end{equation}
On the other hand, for a total energy of electrons with energies $> 1\rm\,TeV$ of  \(W_{\rm e}\sim10^{44}\rm\,erg\) (see Table \ref{tab:fit-param-mwl}), the particle energy density is
\begin{equation}\label{eq:e_energy_density}
w_{\rm e}={W_{\rm e}\over \pi r^2\ell {\cal F}}\simeq 2.7\times 10^{-12}\left({\ell\over1\rm\,pc}\right)^{-1}\rm\,{\cal F}^{-1}\rm\,erg\,cm^{-3}\,,
\end{equation}
where we have taken \(r = 0.63\rm\,pc\) (which corresponds to the spectrum extraction radius of \(7.5'\) at the source distance of
\(290\rm\, pc\)), and $\ell$ is the rather uncertain size of the emitting region along the line of sight. Another rather uncertain parameter, \({\cal F}\), is the filling factor that determines the fraction of the volume filled by relativistic electrons. For large filling factors, \({\cal F}\simeq1\), the cocoon is exclusively occupied by relativistic electrons, and smaller values imply a significant mixing of relativistic plasma and SN ejecta.
The projected size of the cocoon is about \(2{\rm\, pc}\times7{\rm\,pc}\) (which corresponds to angular size of \(0.4^\circ\times1.4^\circ\) at the source distance of \(290\rm\, pc\)), thus it is feasible that \(\ell\gtrsim1{\rm \,pc}\).  Unless $\ell$ is large, $\ell\gg1\rm\,pc$, the TeV particle pressure coincides within a factor of a few with the magnetic field pressure in the cocoon
for \({\cal F}\simeq1\).

Observations of the jet-torus structure in the inner nebula by \textit{Chandra} point to a size of the wind termination shock \(R_{\rm s}\simeq1.3\times10^{17}\rm\,cm\) \citep[i.e. \(30''\) for a distance of \(290\rm\,pc\), where \(30''\) is the projected radius of the X-ray arc in Fig.~2 of][]{2001ApJ...556..380H}. For the pulsar spin-down luminosity of \(\dot{E}\simeq7\times10^{36}\rm\,erg\,s^{-1}\) the Rankine-Hugoniot relations for a weakly magnetised ultrarelativistic pulsar wind \citep{1984ApJ...283..694K} yield a total pressure at the termination shock of
\begin{equation}\label{eq:p_inner}
p_{\rm s}\simeq{\dot{E}\over 6\pi R_{\rm s}^2 c}\simeq7\times10^{-10}\rm\, erg \,cm^{-3}\,.
\end{equation}
On the other hand, the dynamic pressure inside the SNR shell can be obtained under the assumption
that it is in the Sedov-Taylor phase of its evolution by combining Eqs.~27 and~28 from \citet{2018ApJ...860...59K}:  
\begin{eqnarray}
p_{\rm shell} & \simeq & 0.04\, \rho_\textsc{ism} \left({R_\textsc{snr}\over t_\textsc{snr}}\right)^2 = \nonumber  \\
& = &1.4\times10^{-9} \left({n_\textsc{ism}\over 1\,{\rm cm^{-3}}}\right) \left({R_\textsc{snr}\over 15\,{\rm pc}}\right)^2\left({ t_\textsc{snr}\over 10\,{\rm kyr}}\right)^{-2}\rm\, erg \,cm^{-3}\,, \label{eq:p_shell}
\end{eqnarray}
where  \(R_\textsc{snr}\) and \(t_\textsc{snr}\) are the SNR radius and age, and \(n_\textsc{ism}\) is the nucleon number density in the interstellar medium (ISM) surrounding the shell, which is inferred to be $1-2$~cm$^{-3}$ \citep{dubner1998}. 

Equations~\ref{eq:p_inner} and~\ref{eq:p_shell} rely on completely different observational constraints and therefore provide a robust order-of-magnitude estimate of the pressure of $10^{-9}$~erg~cm$^{-3}$.
The pressure estimates by Eqs.~\ref{eq:p_inner} and~\ref{eq:p_shell} significantly exceed the pressure of the TeV particles and magnetic field in the cocoon (Eqs.~\ref{eq:b_energy_density} and \ref{eq:e_energy_density}). This is somewhat surprising, since the MHD flow in PWNe is expected to be subsonic  \citep[see, e.g.][]{KenCor84}, in other words, nearly isobaric.
Therefore, there should be another dominant contribution to the pressure. The energy required to support the pressure in the cocoon can be estimated as
\begin{equation}\label{eq:e_cocoon}
  E_{\rm c}=3V_{\rm c} p\simeq10^{48}\left({\ell\over1\rm\,pc}\right)\rm\,erg\,,
\end{equation}
where we estimated the cocoon volume as
\(V_{\rm c}\simeq 14\times\ell{\rm\, pc^2}\) (which corresponds to angular size
of \(0.4^\circ\times1.4^\circ\) at the source distance of \(290\rm\, pc\)). 

\citet{slane2018} obtained that the size of the Vela X PWN and SNR can be reproduced if one adopts an initial spin-down power of \(\dot{E}_0=10^{39}\rm\,erg\,s^{-1}\) and braking index
\(n_{\rm b}=3\). For these parameters of the pulsar, the total energy injected to the PWN can be estimated as \(3\times10^{49}\rm\,erg\). Although the entire nebula is much bigger than the cocoon and particles may undergo significant energy losses due to radiative and adiabatic cooling, it is possible that relativistic particles could provide the missing pressure.

The pressure deficit cannot be explained by sub-TeV electrons in the cocoon, since \citet{tibaldo2018} found that their spectrum is very hard. Alternatively, the missing pressure could be provided by relativistic electrons in the extended radio nebula. According to \citet{grondin2013} their total energy is about \(\simeq10^{48}\rm\,erg\). These electrons are distributed in a region significantly larger than the cocoon (probably, a factor of approximately ten in volume), thus their contribution is not sufficient to explain completely the pressure deficit either.
Finally,  even mildly relativistic electrons with such energy density could violate the upper limits derived at \(400\rm\,MHz\) on the radio brightness from the Vela X direction \citep{1982A&AS...47....1H,grondin2013}.

Since the radio data do not allow us to constrain a possible contribution to pressure from relativistic ions, they may be considered a candidate to explain the pressure deficit. Models of pulsar winds that include relativistic ions were presented, for example, by \citet{arons1994}. \citet{horns2006} proposed a model of gamma-ray emission from Vela~X in which all the emission is produced by relativistic ions with a total energy of $10^{49}$~erg for protons or $10^{48}$~erg for iron nuclei. We note, however, that this scenario may be problematic when considering the energetics of relativistic ions and leptons over the whole PWN, including the extended radio nebula, as discussed in \citet{dejager2009}. We also note that in this scenario the interpretation of our gamma and X-ray data would require significant modification.

Alternatively, the pressure deficit could be considered as an indication of a significant contribution from non-relativistic ejecta material to the pressure in the cocoon region, that is, by taking a small filling factor \({\cal F}\leq10^{-2}\) in Equation~\ref{eq:e_energy_density}.
This is qualitatively consistent with the results from the hydrodynamic simulations of \citet{slane2018} that show how the impact of the reverse shock also causes a pronounced mixing of the ejecta with young PWN material in the cocoon area \citep[Fig.~13 of][]{slane2018}. 
We note, however, that the study by \citet{slane2018} was performed in the hydrodynamic limit, thus it does not allow robust conclusions regarding the properties of the magnetic field. In the future, this scenario should be revisited in the MHD framework.  

\subsection{Origin of the cocoon and particle transport}
We have considered two scenarios to explain the origin of the relativistic particles in the Vela X cocoon. In the first scenario particles diffuse from the acceleration site, which is conventionally associated with the pulsar wind termination shock, to the cocoon \citep[e.g.][]{hinton2011}. In this case the cocoon represents just a preferred particle leakage path due to some peculiar configuration of the magnetic field \citep[see, e.g. the discussion in ][]{dejager2009}. In the second scenario, particles are quickly advected from the acceleration site to the cocoon by the highly asymmetric reverse shock \citep{blondin2001,slane2018}. We note that  proper modelling of these two processes requires detailed MHD and particle transport simulations, that are beyond the scope of this paper. In the following we compare some of the results obtained from our analysis to general expectations for the two scenarios. 

As shown above, the TeV particle and magnetic field energy densities are comparable in the cocoon, therefore particle diffusion must proceed in the non-linear regime. Nevertheless it is worthwhile to compare the required diffusion coefficient with the Bohm limit:
\begin{equation}
D=\kappa D_{\rm B}\sim10^{25}\kappa \left({E\over 1\rm\,TeV}\right)\left({B\over 6\upmu G}\right)^{-1}\rm\,cm^2\,s^{-1}\,,
\end{equation}
where \(\kappa\) is a dimensionless scaling constant. The characteristic diffusion distance is then 
\begin{equation}
\Delta r\sim\sqrt{D\Delta t} \sim 0.5 \kappa^{1/2} \left({E\over 1\rm\,TeV}\right)^{1/2}\left({B\over 6\upmu G}\right)^{-1/2}\left({\Delta t \over 10^{4}\rm\,yr}\right)^{1/2}\rm\,pc\,,
\end{equation}
which is comparable to the size of the cocoon  for \(E\sim1\rm\,TeV\) if \(\kappa\geq100\). Although the diffusion should proceed much faster than in the Bohm limit,  this value of the diffusion coefficient is still significantly smaller than typical ISM values \citep[e.g.][]{strong2007}, which is expected, to some extent, in media perturbed by an SN explosion or due to non-linear particle transport \citep{evoli2018}. %Thus, the current observational data seem to be  compatible with the diffusion scenario. LT: will leave this for the conclusions.

Even for a weak magnetic field in the cocoon, synchrotron emission should be the dominant cooling channel. Thus, the cooling time can be estimated as
\begin{equation}\label{eq:t_cool}
t_{\rm syn}=3.5\times10^5 \left({E\over 1\rm\,TeV}\right)^{-1}\left({B\over 6\rm\,\upmu G}\right)^{-2} \rm \, yr\,.
\end{equation}
In the diffusion scenario, the particles accelerated in the inner nebula may experience a strong magnetic field. The pressure near the termination shock estimated above corresponds to an equipartition magnetic field of  \(B_{\rm s}\simeq100\rm\,\upmu G\), which for \(100\rm\, TeV\)  electrons yields a cooling time of  \(t_{\rm syn}\sim10\rm\,yr\), which is very short as compared to the age of the source. However, particles may escape quickly from the high-magnetic field strength region (for ballistic propagation ultra-relativstic particles would take $\sim$1~yr to reach the lower magnetic field strength region in pointing 0), and it cannot be excluded that the magnetic field strength is in fact weaker than the equipartition value. To assess if the diffusion scenario is viable the transport of the particles near the termination shock must be studied in detail, which is, however, beyond the scope of this paper \cite[for some studies in this direction, see, e.g.][]{2011ApJ...742...62V,porth2016,2018ApJ...867..141I}.

Alternatively, if the cocoon was created by the reverse shock, then it would correspond to a region that was originally close to the termination shock and which was then swept away. Before the arrival of the reverse shock, the radius of the termination shock was significantly larger than at present, and consequently the magnetic field at the termination shock could be smaller. In this case since the formation of the cocoon proceeds by a hydrodynamic process, the particle transport occurs in an energy independent manner. 
For the observed magnetic field strength of \(B\simeq6\rm\,\upmu G\) and a time of  6000~yr, the age of the cocoon suggested by \citet{slane2018}, a synchrotron cooling feature appears at \(E_{\rm cut}\simeq100\rm\,TeV\), consistent  with our results (Table \ref{tab:fit-param-mwl}). In this case, the reconstructed spectrum below \(100\rm\,TeV\) still corresponds to the acceleration spectrum.  Table~\ref{tab:fit-param-mwl} shows that the values of the power-law index are broadly consistent with the canonical value of \(\alpha\simeq2-2.2\) expected at late times for Fermi acceleration at relativistic shocks \citep[e.g.][]{achterberg2001,sironi2007}, but uncertainties remain large.

\section{Summary and conclusions}\label{sec:summary}

We have combined X-ray data from three pointings of \suz (total observation time of 139~ks) with $\sim 100$~hours of gamma-ray observations with \hess to extract the SEDs in three compact ($\sim$1~pc) regions of the Vela~X cocoon covering distances from the pulsar that range from \(\sim0.3\rm\,pc\) to \(\sim4\rm\,pc\) (for a distance of 290~pc). The gamma-ray spectra are best modelled  by power laws with exponential cutoffs in pointings~1 and 2 (farther from the pulsar). For pointing 0 (closer to the pulsar) the best-fit exponential-cutoff power law has a cutoff energy consistent within statistical uncertainties with the other two regions, although the presence of a cutoff is not statistically significant ($<3\sigma$). The X-ray spectra are all well described by power laws modified by Galactic interstellar absorption. The X-ray spectral properties of regions~1 and 2 are consistent within uncertainties, while region~0, closer to the pulsar, shows a suggestion of harder emission, although uncertainties from the spill-over of emission outside our spectral extraction region are larger.

We have fitted the X-ray and TeV SEDs with a simple radiative model with an electron population producing synchrotron and IC emission. This enabled us to
reconstruct the electron spectrum and magnetic field properties with minimal modelling assumptions. For the electron spectral distribution we adopted a four-parameter family of power laws with sub or super-exponential cutoffs. For the magnetic field we considered two different scenarios: a fixed strength, and a case in which
the magnetic field also has a component with power-law strength distribution that aims to account for the contribution from turbulent fluctuations. The results did not favour the presence of a significant turbulence level: the constant magnetic field scenario with fewer parameters can satisfactorily reproduce the data. The parameters of the electron spectral distribution are consistent in the two cases within the uncertainties, which are particularly large for the exponential cutoff index.

Magnetic field and TeV electrons are in a state close to energy equipartition in all three regions. The pressure of both components ($\sim$10$^{-12}$~erg~cm$^{-3}$) is small compared to what is inferred from the properties of the wind termination shock and the SNR shell ($\sim$10$^{-9}$~erg~cm$^{-3}$). This  points to the existence of another dominant source of pressure. Sub-TeV electrons can hardly explain the pressure deficit completely, but a contribution from relativistic ions cannot be excluded based on the current data. An alternative explanation may be found in a significant contribution from non-relativistic matter due to mixing of PWN and ejecta (small filling factor for relativistic particles), consistent with recent hydrodynamical simulations of the system \citep{slane2018}. 

We compared the electron spectra with general expectation from particle-transport scenarios dominated by either diffusion or advection via the reverse shock. In the first case the diffusion coefficient needs to be 100 times larger than the Bohm limit, but still significantly smaller than typical ISM values. Significant radiative losses expected near the termination shock require to carefully study particle transport in this region in order to assess if the scenario is viable in the light of the measured cutoff energies of $\sim$100~TeV. Conversely, in the reverse-shock scenario the cutoff energies are consistent with radiative cooling in the cocoon, and the spectra below the cutoff, corresponding to the acceleration spectra, are broadly compatible with expectations for Fermi acceleration at relativistic shocks. 

The constraints on magnetic field turbulence and the shape of the particle cutoff are very sensitive to the spectrum in
the hard X-ray band. Thus, they could be improved with future observations in this domain, for example, with \nustar \citep{harrison2013}. They may also benefit from better gamma-ray measurements with CTA \citep{actis2011}.

\begin{acknowledgements}
The support of the Namibian authorities and of the University of Namibia in facilitating the construction and operation of H.E.S.S. is gratefully acknowledged, as is the support by the German Ministry for Education and Research (BMBF), the Max Planck Society, the German Research Foundation (DFG), the Alexander von Humboldt Foundation, the Deutsche Forschungsgemeinschaft, the French Ministry for Research, the CNRS-IN2P3 and the Astroparticle Interdisciplinary Programme of the CNRS, the U.K. Science and Technology Facilities Council (STFC), the IPNP of the Charles University, the Czech Science Foundation, the Polish National Science Centre, the South African Department of Science and Technology and National Research Foundation, the University of Namibia, the National Commission on Research, Science \& Technology of Namibia (NCRST), the Innsbruck University, the Austrian Science Fund (FWF), and the Austrian Federal Ministry for Science, Research and Economy, the University of Adelaide and the Australian Research Council, the Japan Society for the Promotion of Science and by the University of Amsterdam.
We appreciate the excellent work of the technical support staff in Berlin, Durham, Hamburg, Heidelberg, Palaiseau, Paris, Saclay, and in Namibia in the construction and operation of the equipment. This work benefited from services provided by the H.E.S.S. Virtual Organisation, supported by the national resource providers of the EGI Federation.

This research has made use of data obtained from the \suz satellite, a collaborative mission between the space agencies of Japan (JAXA) and the USA (NASA). 

We acknowledge the usage of the following open-source software: aplpy \citep{aplpy}, astropy \citep{astropy2}, corner \citep{corner}, emcee \citep{foreman2013}, matplotlib \citep{hunter2007}, naima \citep{zabalza2015}.

\end{acknowledgements}

\newpage

\bibliographystyle{aa} % style aa.bst
\bibliography{../ref} % your references Yourfile.bib

\newpage

\begin{appendix}

\section{Additional Information from the \hess spectral fitting}\label{app:hessfit}

\begin{table}[!htbp]
\caption{Counting statistics in the spectral extraction region (ON) and background evaluation region (OFF) integrated over the whole energy range $>0.6\rm\,TeV$, and significance of the gamma-ray excess for the three analysis regions.}\label{tab:gammacnt}
\centering
\begin{tabular}{lccc}
\hline\hline
                                        & Pointing 0    & Pointing 1    & Pointing 2\\
\hline
$N_\mathrm{ON}$         & 262           & 724           & 768\\
$N_\mathrm{OFF}$                & 2825          & 3777          & 4560\\
$\alpha$                                & 0.048         & 0.092         & 0.078\\
$N_\mathrm{background}$& 135.9          & 348.0         & 355.1\\
$N_\mathrm{excess}$     & 126.1         & 376.0         & 412.9\\
Significance ($\sigma$) & 9.3                   & 16.6          & 18.0\\
\hline
\end{tabular}
\tablefoot{Parameter $\alpha$ is defined such that the number of background counts expected in the ON region is $N_\mathrm{background} = \alpha \times N_\mathrm{OFF}$ \citep{berge2007}. The significance of the gamma-ray excess is evaluated according to \citet{lima1983}.}
\end{table}

\section{Contributions from infrared fields to inverse-Compton emission from Vela~X}\label{app:irfield}

\begin{figure}
\centering
\includegraphics[width=0.5\textwidth]{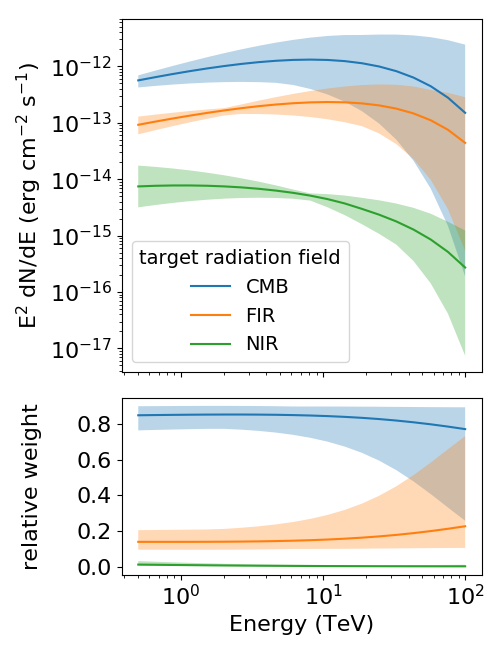}
\caption{Contributions from different target radiation fields to IC emission from a hypothetical source with properties similar to Vela X as observed by \hess The top panel shows the SED from the three individual target radiation fields. The bottom panel shows the relative weight of each component with respect to the total. The three target radiation fields are: the cosmic microwave background (CMB), a far-infrared (FIR) field parametrised as a diluted blackbody with temperature of 30~K and energy density of \(0.2\rm\,eV\,cm^{-3}\), and near-infrared (NIR) field parametrised as a diluted blackbody with temperature of 3000~K and energy density of \(1\rm\,eV\,cm^{-3}\). The parent electron population has a spectrum parametrised by Equation~\ref{eq:espec}. We adopt $A=3\times10^{29}\rm\,eV^{-1}$, and $E_0 = 10\rm\,TeV$. The reference spectra (solid lines in the top panel) correspond to $\alpha=1.75$, $E_\mathrm{cut}=70\rm\,TeV$, and $\beta = 1.2$. The shaded bands show the range spanned by variations of the electron spectrum with $1.4< \alpha < 2.1$, $40\,\mathrm{TeV}<E_\mathrm{cut}<100\,\mathrm{TeV}$, and $0.9< \beta < 1.5$ (cf. Table~\ref{tab:fit-param-mwl}). The assumed distance of the source is \(290\rm\,pc\).}
\label{fig:fieldweight}
\end{figure}

The interpretation of the SED of the Vela~X cocoon has often taken into account the CMB alone as target radiation field for IC scattering of the accelerated electrons \citep[e.g.][]{hess2006velax}. In Figure~\ref{fig:fieldweight} we show the contributions from different target radiation fields that approximate the interstellar radiation field observed near the Sun to the IC emission from a source with spectrum analogous to Vela~X.
Due to the Klein-Nishina suppression IC emission from higher-energy target photons is less important. The NIR component in Figure~\ref{fig:fieldweight} contributes $<2\%$ to the total IC emission at any energies, that is, direct starlight is unimportant. On the other hand, the FIR component in Figure~\ref{fig:fieldweight} accounts for $>10\%$ of the total IC emission at any energies. Moreover, if a cutoff of the electron spectrum at few tens of TeV is present, as in Vela~X, the competition between the Klein-Nishina suppression and the energy of IC photons upscattered from higher-energy target photons being higher can make IC emission from the FIR field prevail over that from the CMB. This shows that accounting for thermal emission from dust can be important, depending on the exact spectrum of the electron population \cite[see also][]{1997MNRAS.291..162A}.

In our analysis we adopted the FIR model by \citet{popescu2017}. This should be considered as a large-scale average, and therefore close to a lower limit on the intensities of the FIR fields at the position of our source. We verified that individual IR sources in the IRAS Catalogue \citep{iras1988} contribute negligible fluxes even under the hypothesis that they all lie at the same distance from the Earth as Vela~X. However, the diffuse fluxes may be larger due to local interstellar structures not accounted for in the model by \citet{popescu2017}. We have assessed the impact on the results of an increase of a factor of two of the FIR energy density on the fit parameters. The results are reported in Table~\ref{tab:fit-param-mwl-highFIR}.
   \begin{table}[!htbp]
      \caption{Best-fit parameters for the radiative model of the X-ray and gamma-ray spectrum of the Vela X cocoon for the case of increased FIR intensities in the three regions shown in Figure~\ref{fig:maps}.}
         \label{tab:fit-param-mwl-highFIR} 
         \centering
         \begin{tabular}{lcccccc}
            \hline\hline
            Pointing            &  $W_e$ ($> 1\rm\,TeV$)\tablefootmark{a}                               & $\alpha$\tablefootmark{b}               & $E_\mathrm{cut}$\tablefootmark{b} & $\beta$\tablefootmark{b} & $B$\tablefootmark{c} & BIC\tablefootmark{d} \\
                                & ($10^{44}$ erg)       &                               & (PeV) & & ($\upmu$G)\\
            \hline
            0                   & $0.6_{-0.2}^{+0.5}$ & $2.2_{-0.6}^{+0.4}$ & $0.3_{-0.2}^{+2.0}$ & $2.1_{-1.6}^{+2.6}$ & $9_{-2}^{+4}$ & 14.8\\
            1                   & $1.4_{-0.4}^{+0.6}$ & $1.8_{-0.5}^{+0.4}$ & $0.05_{-0.03}^{+0.06}$ & $0.9_{-0.3}^{+0.6}$ & $7.3_{-0.9}^{+1.1}$ & 18.1\\
            2                   & $1.7_{-0.5}^{+0.6}$ & $2.0_{-0.4}^{+0.3}$ & $0.12_{-0.05}^{+0.03}$ & $2.3_{-1.0}^{+2.2}$ & $5.8_{-0.8}^{+0.9}$  & 17.3 \\
            \hline
         \end{tabular}
         \tablefoot{Median values from the MCMC scan, with lower and upper uncertainties based on the 16th and 84th percentiles of the posterior distribution.\\
\tablefoottext{a}{Total electron energy for particle energies $>1\rm\,TeV$. We note that the solid angle subtended by region~0 is 41\% of that in regions~1 and~2.}\\
\tablefoottext{b}{Parameters of the electron spectrum as defined in Equation~\ref{eq:espec}.}\\
\tablefoottext{c}{Strength of the magnetic field.}\\
\tablefoottext{d}{Bayesian information criterion, i.e. $k \cdot \ln n + \chi^2$, where $k$ is the number of parameters estimated from the model and $n$ is the number of data points.}}
   \end{table}
The comparison with Table~\ref{tab:fit-param-mwl} shows that the effect is smaller than other uncertainties in the parameters  \cite[in agreement with the general conclusion by][]{1997MNRAS.291..162A}. 

\section{Probability density distributions of radiative model parameters fit to the multi-wavelength data}\label{app:corner}

\begin{figure*}[!htbp]
\centering
\includegraphics[width=1\textwidth]{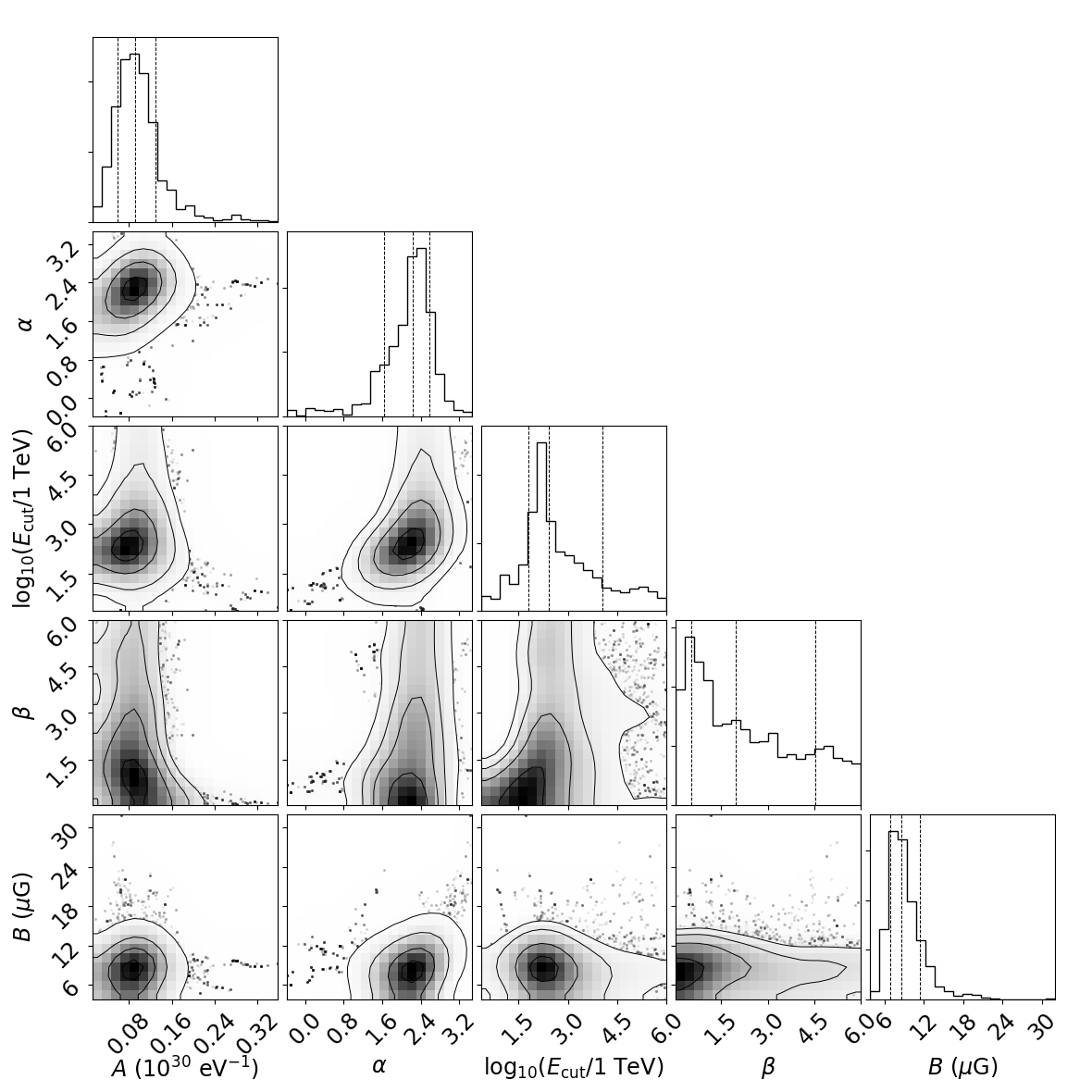}
\caption{One- and two-dimensional projections of the posterior probability density distributions of the parameters for the radiative model of the X-ray and gamma-ray spectrum of the Vela X cocoon for Pointing~0. The parameters of the electron spectrum are defined in Equation~\ref{eq:espec}, and $B$ is the strength of the magnetic field. The lines overlaid on the one-dimensional projections are the 16th, 50th and 84th percentiles of the distributions. The contours overlaid to the two-dimensional projections correspond to $1\sigma$, $2\sigma$, $3\sigma$, and $4\sigma$ probability decrease with respect to the maximum.}
\label{fig:corner0}
\end{figure*}
\begin{figure*}
\centering
\includegraphics[width=1\textwidth]{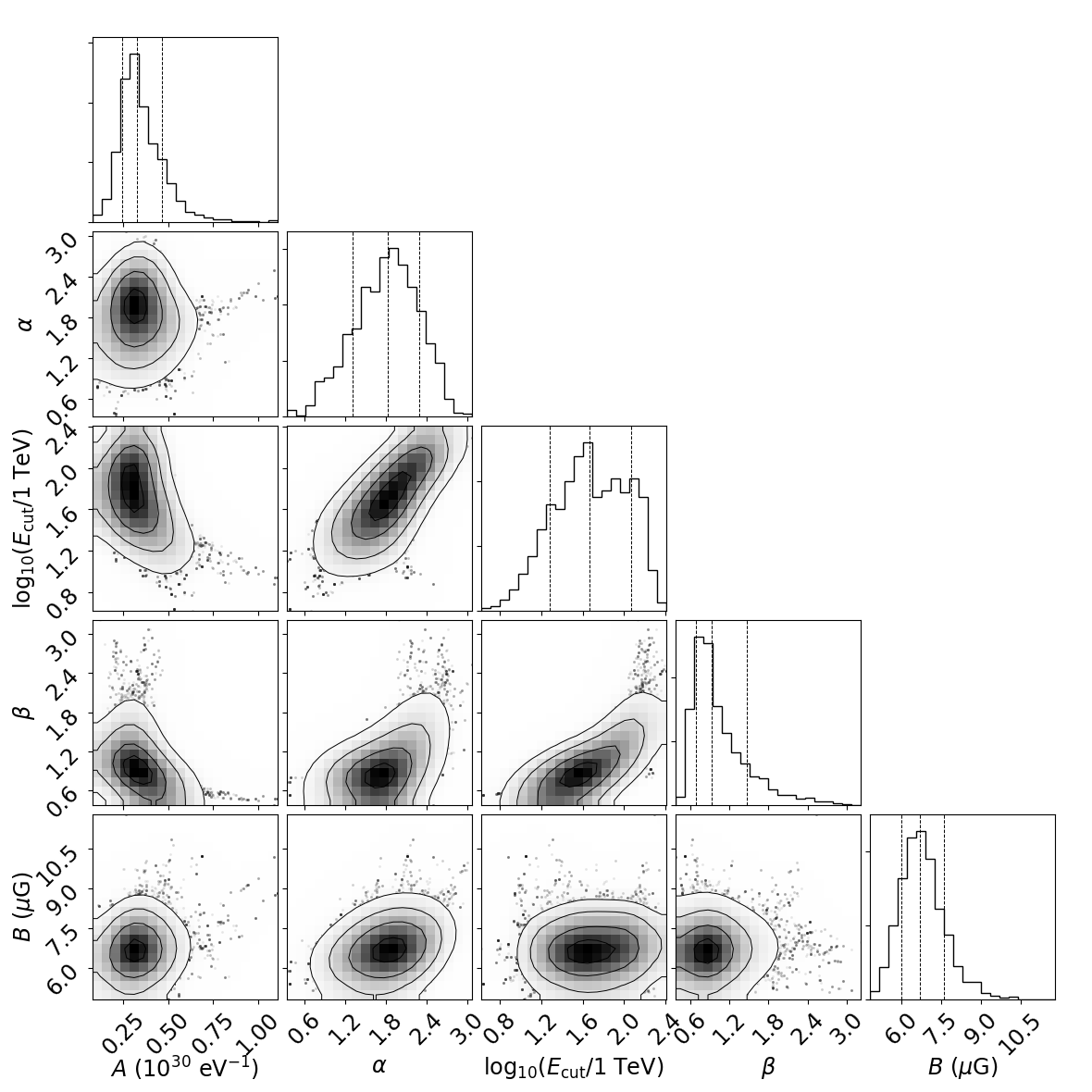}
\caption{One- and two-dimensional projections of the posterior probability density distributions of the parameters for the radiative model of the X-ray and gamma-ray spectrum of the Vela X cocoon for Pointing~1. See Figure~\ref{fig:corner0} for further detail.}
\label{fig:corner1}
\end{figure*}
\begin{figure*}
\centering
\includegraphics[width=1\textwidth]{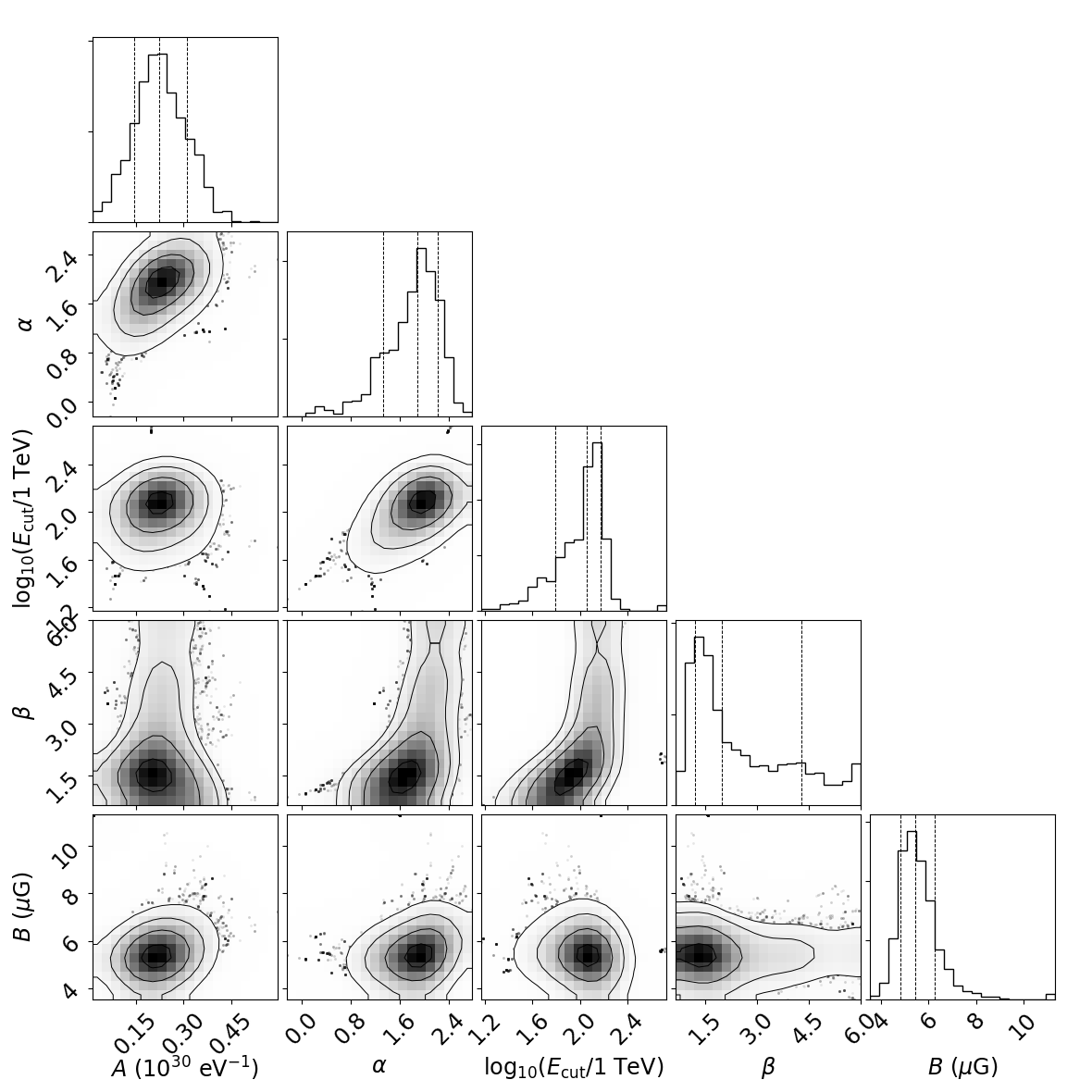}
\caption{One- and two-dimensional projections of the posterior probability density distributions of the parameters for the radiative model of the X-ray and gamma-ray spectrum of the Vela X cocoon for Pointing~2. See Figure~\ref{fig:corner0} for further detail.}
\label{fig:corner2}
\end{figure*}

\end{appendix}

\end{document}